\documentclass[12pt]{article}

\usepackage{epsf,epsfig}
\usepackage{a4wide}

\def\slash#1{#1 \hskip-0.45em /}
\def\Slash#1{#1 \hskip-0.59em /}

\def\beq{\begin{eqnarray}}
\def\eeq{\end{eqnarray}}

\def\be{\begin{equation}}
\def\ee{\end{equation}}

\def\np{n_+}
\def\nm{n_-}

\newcommand{\SCETI}{\mbox{SCET(hc,c,s)}}
\newcommand{\SCETII}{\mbox{SCET(c,s)}}

\begin{document}
\thispagestyle{empty}

\begin{flushright}
  PITHA 03/11\\
  CERN-TH/2003-286\\
  hep-ph/0311335\\
  November 25, 2003
\end{flushright}

\vspace{\baselineskip}

\begin{center}
\vspace{0.5\baselineskip}
\textbf{\Large 
Factorization of heavy-to-light form factors\\[0.5em]
in soft-collinear effective theory}\\
\vspace{3\baselineskip}
{\sc M. Beneke$^a$ and Th.~Feldmann$^b$}\\

\vspace{0.7cm}
{\sl ${}^a$Institut f\"ur Theoretische Physik E, RWTH Aachen\\
D--52056 Aachen, Germany\\
\vspace{0.3cm}
${}^b$CERN, Theory Division, CH-1211 Geneva, Switzerland}\\
\vspace{3\baselineskip}

\vspace*{0.5cm}
\textbf{Abstract}\\
\vspace{1\baselineskip}
\parbox{0.9\textwidth}{
Heavy-to-light transition form factors at large recoil energy of the 
light meson have been conjectured to obey a factorization formula, 
where the set of form factors is reduced to a smaller number of 
universal form factors up to hard-scattering corrections. In this 
paper we extend our previous investigation of heavy-to-light 
currents in soft-collinear effective theory to final states 
with invariant mass $\Lambda^2$ as is appropriate to exclusive 
$B$ meson decays. The effective theory contains soft
modes and two collinear modes with virtualities of order 
$m_b\Lambda$ (`hard-collinear') and $\Lambda^2$. Integrating out
the hard-collinear modes results in the hard spectator-scattering
contributions to exclusive $B$ decays.  We discuss the 
representation of heavy-to-light currents in the effective theory 
after integrating out the hard-collinear scale, and show that 
the previously conjectured factorization formula is valid to all 
orders in perturbation theory. The naive factorization of matrix 
elements in the effective theory into collinear and soft matrix
elements may be invalidated by divergences in convolution integrals. 
In the factorization proof we circumvent the explicit regularization
of endpoint divergences by a definition of the universal form factors 
that includes hard-collinear, collinear and soft effects. }
\end{center}

\newpage
\setcounter{page}{1}

\section{Introduction}
\label{sec:intro}

The purpose of this paper is to develop further the theory of
exclusive $B$ decays to light energetic mesons. We are specifically interested
in ``hard-spectator scattering'', i.e. scattering
mechanisms that involve the light antiquark in the $\bar B$ meson. The
soft external quark line is one of the crucial
differences to the standard situation of hard 
exclusive processes \cite{Lepage:1980fj} 
involving only light hadrons, where all external lines carry large
momentum. A consequence of this difference is that exclusive $B$ meson
decays involve two hard scales, $m_b^2$ and $m_b\Lambda$, where
$\Lambda$ is of order of the strong interaction scale. 

Hard-spectator scattering is an important ingredient in QCD
factorization for non-leptonic $B$ decays to charmless final 
states \cite{Beneke:1999br,Beneke:2000ry},
and is even more important in the so-called PQCD 
approach \cite{Keum:2000wi}. A better
understanding of spectator interactions is needed to justify the
factorization hypotheses of the two approaches to all orders in
perturbation theory and to leading order in the heavy-quark
limit. However, even apparently simpler processes such as the
semi-leptonic decay $B\to \pi l\nu$ at large momentum transfer to the
pion are currently not completely understood. The one exception is 
$B\to \gamma l\nu$, which has received much attention recently 
\cite{Korchemsky:1999qb,Descotes-Genon:2002mw,Lunghi:2002ju,Bosch:2003fc}.
A factorization formula of the schematic form $F=T\star \phi_B$, where 
the star-product denotes convolution of a hard-scattering kernel with
the $B$ meson light-cone distribution amplitude,  has now
been shown to be valid to all orders in perturbation theory 
\cite{Lunghi:2002ju,Bosch:2003fc}.
In this paper we consider $B\to \pi l\nu$ (heavy-to-light form
factors) at large pion energy. A summary of the results has 
already been given in \cite{Beneke:2003xr}.

A straightforward extension of the results for $B\to\gamma l\nu$ 
decay to the $B\to \pi$ form 
factors relevant to semi-leptonic decays fails. If one writes 
the form factors as $\phi_\pi\star T\star\phi_B$ in analogy to the
$\pi\to\pi$ transition form factor at large momentum transfer, 
one finds that the convolution integrals do not converge at their 
endpoints. In other words, the form factors receive leading
contributions from momentum configurations where some partons in the
pion appear to have small momentum \cite{Szczepaniak:1990dt,Chernyak:ag}. 
In \cite{Beneke:2000wa} the factorization formula 
\begin{equation}
\label{ff1}
F_i = C_i \,\xi_\pi + \phi_B\star T_i\star \phi_\pi
\end{equation}
has been conjectured for the three Lorentz invariant 
$B\to \pi$ form factors and shown to be valid at order 
$\alpha_s$. The additional term $C_i\,\xi_\pi$ involves a short-distance
coefficient and a single ``soft form factor'' $\xi_\pi$, which obeys the
large-recoil symmetries \cite{Charles:1998dr}. 
Factorization for the  
$B\to \pi$ form factors is more complicated than for both, the $\pi\to\pi$
and $B\to D$ form factors. At large momentum transfer soft
interactions cancel in the $\pi\to\pi$ transition at leading
power. The remaining hard and collinear effects are factored into
convolutions as in the second term on the right-hand side of
(\ref{ff1}). When both mesons are heavy, such as in $B\to D$,
collinear effects are irrelevant. The remaining hard and soft
interactions factor into a short-distance coefficient and the
Isgur-Wise form factor \cite{Isgur:1990ed} 
similar to the first term on the right-hand
side of (\ref{ff1}). The $B\to \pi$ form factor, however, involves
hard, collinear and soft effects. Furthermore, due to the presence of
several scales one must distinguish short- and long-distance collinear
effects, as we discuss in more detail below. 

A separation of all these
effects and an operator definition of the various short- and
long-distance quantities in (\ref{ff1}) to all orders in $\alpha_s$
and to leading order in $1/m_b$ has not yet been achieved. A
complication was pointed out in \cite{Bauer:2002aj}, where
it was shown that superficially sub-leading interactions in $1/m_b$ 
contribute to the $B\to\pi$ form factors at leading power. Some of the 
previous arguments \cite{Bauer:2000yr,BCDF} to justify or
extend some aspects of (\ref{ff1}) must therefore be revised. 
The form factors have been reconsidered in \cite{Bauer:2002aj} 
in the framework of soft-collinear effective field theory (SCET) 
\cite{Bauer:2000yr,Bauer:2001yt} and the structure of the formula 
(\ref{ff1}) was seen to emerge. However, in  \cite{Bauer:2002aj} a
technical definition of ``factorizable'' and ``non-factorizable'' 
terms has been adopted that does not correspond to the usual 
notions, so that the issues of endpoint singularities and convergence 
of convolution integrals could not be clarified. Below 
we extend SCET in the position-space 
formulation \cite{BCDF,Beneke:2002ni} to cover the case of exclusive
decays, where the external collinear lines have invariant mass 
of order $\Lambda^2$ as appropriate for exclusive decays. 
To obtain the factorization formula (\ref{ff1}) we 
match SCET to an effective theory from which short-distance collinear modes
with virtuality $m_b\Lambda$ are removed. This point was first
addressed in \cite{Hill:2002vw} and the formalism has been worked out
to the extent that a factorization theorem for $B\to\gamma l\nu$ was 
established \cite{Bosch:2003fc}. 

\begin{table}
\centerline{\parbox{14cm}{\caption{\label{tab1}\small
Terminology for the various momentum modes relevant to exclusive $B$
decays. The momentum components are given as $(\np p,p_\perp,\nm p)$, 
but mass dimension has to be restored by multiplying appropriate
factors of $m_b$. Two different terminologies for the same momentum
modes are used in the literature. In physical units $\lambda$ is of order 
$(\Lambda/m_b)^{1/2}$, where $\Lambda$ is the strong-interaction 
scale.}}}
\vspace{0.1cm}
\begin{center}
\begin{tabular}{||c|l|l||}
\hline\hline
&& \\[-0.7em]
Momentum scaling & Terminology I & Terminology II
\\[0.3em] 
& {\small (\cite{Bauer:2000yr},\cite{BCDF})} & 
{\small (\cite{Hill:2002vw}, this work)} \\[0.3em]
\hline\hline
&& \\[-0.7em]
$(1,1,1)$ & hard &   hard\\[0.3em]
\hline
&&\\[-0.7em]
$(\lambda,\lambda,\lambda)$ & soft & semi-hard \\[0.3em]
$(1,\lambda,\lambda^2)$ & collinear & hard-collinear \\[0.3em]
\hline
&& \\[-0.7em]
$(\lambda^2,\lambda^2,\lambda^2)$  & ultrasoft    & soft \\[0.3em]
$(1,\lambda^2,\lambda^4)$  & ultracollinear & collinear \\[0.3em]
\hline\hline
\end{tabular}
\end{center}
\end{table}

The momentum modes relevant to the factorization of form factors 
$\langle M(p')|\bar q\hspace*{0.04cm} 
\Gamma b|\bar B(p)\rangle$, where $M$ is a light
meson or a photon with momentum of order $m_b$, 
and $\Gamma$ is a Dirac matrix, 
are summarized in Table~\ref{tab1}. In general we decompose a momentum
as 
\begin{equation}
  p^\mu = (n_+p) \, \frac{n_-^\mu}{2} + p_\perp^\mu + (n_-p)
  \frac{n_+^\mu}{2},
\end{equation}
where $n_\pm^\mu$ are two light-like vectors, $n_+^2=n_-^2=0$ with
$\np\nm = 2$. The reference directions  $n_\pm$ are 
chosen such that the energetic
massless external lines have $\np p$ of order $m_b$. 
As indicated in Table~\ref{tab1} two different
terminologies have been used in the literature which has been the
cause of some confusion. Since in this paper we will construct 
an effective theory for modes with virtuality
$\Lambda^2$ only, we will use the second terminology. The
effective theory then contains soft and collinear modes in agreement
with the standard QCD terminology. For power counting we define the
scaling parameter $\lambda$ to be of order
$(\Lambda/m_b)^{1/2}$. This differs from the convention in  
\cite{Hill:2002vw} where $\lambda$ is of order $\Lambda/m_b$.   

The existence of the various modes follows from the assumption that
the external momenta of scattering amplitudes for exclusive $B$ decays at
large momentum transfer are soft or collinear.\footnote{For the heavy
  quark line we write $p_b=m_b v+r$, where $v$ is the $B$ meson
  velocity. The statement that the external heavy quark is soft refers to the
  fact that after this decomposition the residual momentum $r$ is
  soft.} 
One finds the three characteristic virtualities $m_b^2$, 
$m_b\Lambda$ and $\Lambda^2$ by combining external momenta. For
instance, $m_b^2$ is obtained by adding and squaring a heavy quark and
a collinear momentum, or by squaring the heavy quark
momentum. The intermediate virtuality is typical for
interactions of collinear gluons or light quarks with 
soft gluons or light quarks, while 
$\Lambda^2$ arises in the self-interactions of collinear or soft
modes. 

Soft-collinear effective theory as defined in 
\cite{Bauer:2000yr,BCDF} is the effective theory obtained after
integrating out hard modes of virtuality $m_b^2$. This theory still
contains two types of soft modes, called ``semi-hard'' (virtuality of 
order $m_b\Lambda$) and ``soft''. The semi-hard
modes can be integrated out perturbatively, but it appears that
semi-hard loop integrals always vanish in dimensional regularization
\cite{BCDF}, so semi-hard modes can be ignored in practice. 
The theory also contains two types of collinear modes, called
``hard-collinear'' and ``collinear'' according to their
virtuality. Although each one of these two has been discussed in
previous applications of SCET, the simultaneous presence of two
distinct collinear modes has not been considered in much detail 
up to now. The reason
for this is that previous applications of SCET to semi-inclusive 
$B$ decay, such as $B\to X_s\gamma$ near maximal photon energy 
\cite{Bauer:2001yt}, and 
to $B\to\gamma l\nu$ (at leading order in $1/m_b$) 
are sensitive only to hard-collinear modes 
\cite{Lunghi:2002ju,Bosch:2003fc}. One can therefore match
SCET directly to the standard heavy quark effective theory, which
contains only soft modes. On the other hand, in  
the exclusive decay $B\to D\pi$ \cite{Beneke:2000ry,Bauer:2001cu} 
or hard exclusive scattering of light hadrons \cite{Bauer:2002nz} the 
effects of hard-collinear and soft modes cancel almost trivially at 
leading power in the power expansion. The effective theory at leading
power can then be formulated entirely in terms of collinear modes. 

The outline of this paper is as follows: in Section~\ref{sec:toy} we 
study a scalar integral, which would be relevant to $B\to\gamma l\nu$
decay at sub-leading order in $1/m_b$, using the method of 
expanding by regions 
\cite{Beneke:1998zp}. We demonstrate with this example that separate
contributions from hard-collinear and collinear loop momentum must be
included to reproduce the integral in full ``QCD''. We shall also find
that the collinear and soft contributions are not well-defined
individually in dimensional regularization. The interpretation of
these additional divergences provides an important clue to the problem
of endpoint divergences. In the context of effective field theory 
the additional divergences show that the matrix elements in the 
effective theory of soft and collinear fields do not factorize
(naively) into a product of soft and collinear matrix elements as 
one might have concluded from the decoupling of soft and collinear
fields in the Lagrangian. 

In Section~\ref{sec:tree} we turn our attention to the representation 
of the heavy-to-light current in the effective theory with the
hard-collinear scale removed. We integrate out the hard-collinear modes in
tree graphs by solving the classical field equations for the hard-collinear
quark and gluon fields. Despite the complex branchings of the relevant
trees, the solution can be found by choosing a special
gauge for the calculation and reconstructing the complete result 
through gauge invariance. The first term with a non-vanishing 
$\langle \pi|\ldots|\bar B\rangle$ matrix element in the expansion of 
the current is $\lambda^3$ suppressed, which explains the
$1/m_b^{3/2}$ suppression of heavy-to-light form factors at large
recoil. The calculation also shows that the effective operator is 
highly non-local, implying convolutions in two light-like directions. 
The convolutions are divergent, as expected, but we also find 
that quark-antiquark-gluon amplitudes in the $B$ meson and in the 
light meson
contribute at leading power, which is a new feature of heavy-to-light 
transitions. At the end of this section we briefly discuss 
hard-collinear quantum corrections to determine the general form of 
operators and short-distance
kernels in the effective theory of soft and collinear modes.

The existence of divergent convolutions signals that heavy-to-light 
form factors do not factorize straightforwardly. In 
Section~\ref{sec:formfactor} we return to the factorization 
formula (\ref{ff1}), and show that it is indeed valid 
to all orders in the strong coupling and to leading 
power in $1/m_b$. We shall define the universal form factor $\xi_\pi$
as a matrix element in SCET before integrating out hard-collinear 
effects. We then show that the terms not contained in this definition 
factorize into convolutions of light-cone distribution amplitudes 
with convergent integrals after integrating out hard-collinear 
modes. The proof of convergence relies on power counting, boost
invariance,  and 
the correspondence of collinear and soft endpoint divergences 
through soft-collinear factorization. 
We conclude in Section~\ref{sec:conclude}. 


\section{The scalar ``photon'' vertex}
\label{sec:toy}

The purpose of this section is to demonstrate by the example of a
specific Feynman integral that the distinction of
hard-collinear and collinear modes has a technical meaning. We shall
also see how the factorization of collinear and soft modes introduces 
``endpoint singularities'' in the longitudinal integrations, and how 
the singularities are related to cancel in the sum of all
terms. Finally, we sketch how the diagrammatic result would be
interpreted in the context of effective field theory. 

\begin{figure}[t]
   \vspace{-3.5cm}
   \epsfysize=27cm
   \epsfxsize=18cm
   \centerline{\epsffile{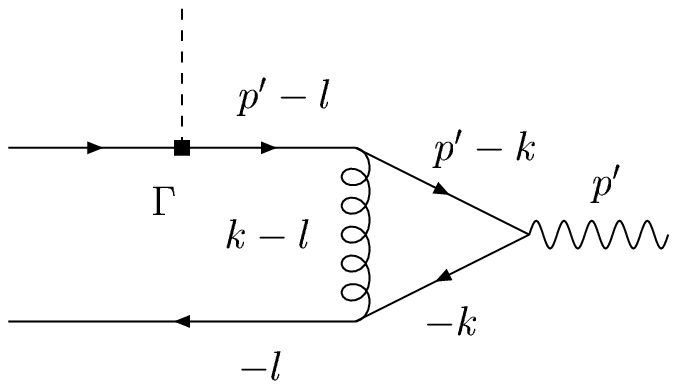}}
   \vspace*{-19.4cm}
\centerline{\parbox{14cm}{\caption{\label{fig:vertex}\small  
Photon vertex correction to $\bar B\to\gamma l\nu$. $\Gamma$ denotes
the weak $b\to u$ decay vertex. We consider the corresponding vertex 
integral with all lines simplified to scalar propagators and all
vertex factors set to 1.}}}
\end{figure}

We consider the scalar integral 
\begin{equation}
\label{defI}
I = \int [dk] \,\frac{1}{[(k-l)^2] [k^2-m^2][(p'-k)^2-m^2]},
\end{equation}
which occurs as a one-particle-irreducible subgraph in the correction
to the radiative decay $\bar B\to\gamma l\nu$ shown in 
Figure~\ref{fig:vertex}. We keep a ``light quark'' mass $m$, which
will take the role of $\Lambda$ as the infrared scale of our problem. 
We define the integration measure as
\begin{equation}
[dk]\equiv \frac{(4\pi)^2}{i}\left(\frac{\mu^2 e^{\gamma_E}}{4 \pi}
\right)^{\!\epsilon} \frac{d^d k}{(2\pi)^d} = 
\mu^{2\epsilon} e^{\epsilon\gamma_E}\,\frac{d^dk}{i\pi^{d/2}}
\qquad (d=4-2\epsilon).
\end{equation}
The integral $I$ is ultraviolet and infrared finite, but 
dimensional regularization will be needed to construct the expansion. 
The $+i\epsilon$ prescription for the propagators is understood.

The external momenta of the vertex subgraph are: a collinear ``photon''
momentum $p'=(\np p',p_\perp^\prime,\nm p')\sim (1,0,0)$ with $p'^{\,2}=0$;
a soft ``light quark'' momentum 
$l\sim (\lambda^2,\lambda^2,\lambda^2)$ with $l^2=m^2$; a 
hard-collinear ``light quark'' momentum $p^\prime-l\sim
(1,\lambda^2,\lambda^2)$ 
with virtuality $\lambda^2$. The two invariants are $2 p'\cdot l\sim
\lambda^2$ and $m^2\sim\lambda^4$, so $I$ must be a function of the 
small dimensionless ratio $m^2/(2p'\cdot l)$. A straightforward 
calculation gives 
\begin{equation}
\label{exactI}
I=\frac{1}{2p'\cdot l} \left(\mbox{Li}_2\left(-\frac{2p'\cdot l}{m^2}
\right)-\frac{\pi^2}{6}\right) 
= -\frac{1}{2p'\cdot l} \left\{\frac{1}{2}\,\ln^2\frac{m^2}{2p'\cdot l} 
+\frac{\pi^2}{3}+\ldots\right\},
\end{equation} 
where after the second equality we neglected higher-order terms in 
$m^2/(2p'\cdot l)$. Below we shall reproduce the first term in the 
expansion by identifying the relevant momentum configurations. 

\subsection{Expansion by momentum regions}

We construct the expansion of $I$ by identifying the momentum
configurations that give non-vanishing contributions to the integral
in dimensional regularization and by expanding the integrand in each
region \cite{Beneke:1998zp}. This method, originally developed for a
non-relativistic expansion and the construction of non-relativistic
effective theory, has also been applied to
integrals with collinear external lines
\cite{Smirnov:1998vk}. 
Here we need a variant for
integrals with collinear and soft external lines. 

To find the relevant momentum regions we first assume that the loop
momentum scales as $k\sim (\lambda^n,\lambda^n,\lambda^n)$ for some 
$n$ and expand the integrand accordingly. For instance, if $k$ is
hard, $n=0$, we find integrals of the type 
\begin{equation}
\int [dk] \,\frac{1}{[k^2]^a [-2p'\cdot k]^b}\times\mbox{polynomial},
\end{equation}
which vanish in dimensional regularization, since the only possible 
invariant $p'^{\,2}=0$. This is not surprising, because there
is no external invariant of order 1. Proceeding for different $n$, the
result is that only the soft momentum region, $n=2$, contributes. 
The corresponding integral will be calculated below.

Since there are external lines with large momentum and small
virtuality, we should also consider loop momentum configurations,
where $\np k$ is the largest component. That is, we take 
$\np k\sim \lambda^n$ and $k^2\sim \lambda^{2 m}$ with $m<n$, expand
the integrand, and determine the integrals that do not
vanish.\footnote{Some integrals vanish independent of any regularization,
  because all poles lie in one of the complex half-planes. Other
  integrals vanish only, because we assume a regularization that does
  not introduce an additional scale into the integral.} The result is
two non-vanishing contributions, one from $\np k\sim 1$ and $k^2\sim
\lambda^2$, which we identify as hard-collinear, and the other from 
$\np k\sim 1$ and $k^2\sim\lambda^4$, which we call collinear, see 
Table~\ref{tab1}. Regions with $k^2<\lambda^4$ do not appear due to
the internal masses $m^2\sim \lambda^4$.  
We will now verify (at leading order) that the sum
of the three regions constructs the expansion of our integral $I$. 

\paragraph{\it The hard-collinear region.} 
Expanding the propagators systematically the leading hard-collinear
integral is
\begin{eqnarray}
\label{hcI}
I_{hc} &=&  \int [dk] \,\frac{1}{[k^2-\np k\nm l] [k^2][k^2-\np
  p^\prime \nm k]} 
\nonumber\\[0.2cm]
&=& -\frac{1}{2p'\cdot l} \left\{
\frac{1}{\epsilon^2}-\frac{1}{\epsilon}\,\ln\frac{2p'\cdot l}{\mu^2}+
\frac{1}{2}\,\ln^2\frac{2p'\cdot l}{\mu^2}-\frac{\pi^2}{12}\right\}.
\end{eqnarray} 
The expansion has rendered the integral infrared divergent. If we 
perform the $\nm k$ integration by contour methods, the $k_\perp$
integral is divergent for $k_\perp\to 0$ (physically, 
$k_\perp\ll \lambda$) for any $\np k$, but the 
$\np k$ integral converges at fixed $k_\perp$. The double pole
originates from $n_\pm  k\to 0$, $k_\perp\to 0$ simultaneously.

\paragraph{\it The collinear region.} 
In this region 
the ``light quark'' propagators with momenta $p'-k$ and $-k$ are
collinear and have virtuality of order $\lambda^4$. The ``gluon''
propagator is hard-collinear with virtuality $\lambda^2$. One finds
that the collinear and soft integrals are not well-defined
separately in dimensional regularization. This also occurred in 
previous applications of the method of expansion by regions to 
collinear integrals \cite{Smirnov:1998vk}, and is related
to the fact the dimensional regulator is attached to the transverse
momentum components. If additional divergences arise from the $\np k$
or $\nm k$ integrations, they may not be regularized. As in 
\cite{Smirnov:1998vk} we introduce an additional
``analytic'' regularization by substituting
\begin{equation}
\frac{1}{[(k-l)^2]}\to\frac{[-\nu^2]^\delta}{[(k-l)^2]^{1+\delta}},
\end{equation}
where $\nu$ is a parameter with mass dimension one. The leading
collinear integral is 
\begin{equation}
\label{collintdef}
I_c = \int [dk] \,\frac{[-\nu^2]^\delta}{[-\np k\nm l]^{1+\delta} 
[k^2-m^2][k^2-m^2-2p'\cdot k]}.
\end{equation}

The integral can be done with standard methods, but it will be useful
to obtain an intermediate result, where only 
the $\nm k$ integration is performed. The variable $\np p'$ is related
to the energy of the ``photon'', so $\np p'>0$. We then close the
contour in the lower half plane and pick up the pole at 
$(-k_\perp^2+m^2-i\epsilon)/\np k$ for $0<\np k<\np p'$. (In 
our convention $k_\perp^2$ is negative.) This gives
\begin{eqnarray}
\label{cI}
I_c &=&  -\frac{1}{2p'\cdot l} \,\left(\frac{\nu^2}{2p'\cdot l}
\right)^\delta \int_0^{\np p'} \frac{d\np k}{\np k}\left(
\frac{\np p'}{\np k}\right)^\delta \,
\left(\mu^2 e^{\gamma_E}\right)^\epsilon
\int\frac{d^{d-2}k_\perp}{\pi^{d/2-1}}\,\frac{-1}{k_\perp^2-m^2}
\nonumber\\[0.2cm]
&=& -\frac{1}{2p'\cdot l} \left(-\frac{1}{\delta}+
\ln\frac{2 p'\cdot l}{\nu^2}\right)\left(\frac{1}{\epsilon}-
\ln\frac{m^2}{\mu^2}\right).
\end{eqnarray}
The pole at $\epsilon=0$ comes from the collinear singularity 
$k_\perp\to\infty$ (for any $\np k$) in the
transverse momentum integral. 
The additional singularity at $\delta=0$
is an ``endpoint divergence'', which arises from a singularity 
at $\np k\to 0$ for any transverse momentum. This does not correspond
to any singularity of the hard-collinear integral. Since $\nm k$
becomes large compared to $\lambda^4$, when $\np k$ becomes small, 
the endpoint singularity is related to a momentum configuration, where
the ``quark'' with momentum $k$ becomes soft. In particular, since 
the endpoint divergence occurs for any $k_\perp\sim\lambda^2$ it
must be cancelled by a momentum region with $k_\perp\sim \lambda^2$. 

Also note that the integral depends non-analytically on the soft 
external momentum component $\nm l$. This is surprising, since we
would have expected factorization of soft and collinear modes, so that
the collinear integrals could depend only analytically on $\nm
l$, and the soft integrals could depend only analytically on $\np
p'$. Indeed, this would be the case in dimensional regularization, 
where the factor $1/\nm l$ in (\ref{collintdef}) could be pulled
out of the integral. However, the integral is not well-defined in
dimensional regularization. The breakdown of naive collinear-soft
factorization is hence a consequence of the need to introduce a
different regularization, here chosen as analytic.

\paragraph{\it The soft region.} In this region the ``gluon''
propagator and the ``light quark'' propagator with momentum $-k$ are
soft and have virtuality of order $\lambda^4$. The ``light quark''
with momentum $p'-k$ is hard-collinear with virtuality
$\lambda^2$. The leading soft integral is  
\begin{equation}
I_s = \int [dk] \,\frac{[-\nu^2]^\delta}{[(k-l)^2]^{1+\delta} 
[k^2-m^2][-\np p' \nm k]}.
\end{equation}
Here we perform the $\np k$ integral first. Assuming $\nm l>0$, we 
close the contour in the lower half plane and pick up the pole at 
$(-k_\perp^2+m^2-i\epsilon)/\nm k$ for $0<\nm k<\nm l$. This gives 
\begin{eqnarray}
\label{sI}
I_s &=&  -\frac{1}{2p'\cdot l}\, 
\int_0^{\nm l} \frac{d\nm k}{\nm k}\left(
\frac{\nm l}{\nm k}\right)^{-\delta} \,
\left(\mu^2 e^{\gamma_E}\right)^\epsilon
\int\frac{d^{d-2}k_\perp}{\pi^{d/2-1}}\,\frac{\nu^{2\delta}}
{(m^2 \left[1-\frac{\nm k}{\nm l}\right]^2-k_\perp^2)^{1+\delta}}
\nonumber\\[0.2cm]
&=& -\frac{1}{2p'\cdot l} \left(\frac{\mu^2 e^{\gamma_E}}{m^2}
\right)^{\!\epsilon} 
\Gamma(\epsilon) \left(\frac{m^2}{\nu^2}\right)^{-\delta}\,
\frac{1}{\delta}\,\frac{\Gamma(\delta+\epsilon)
  \Gamma(1-2\delta-2\epsilon)}{\Gamma(\epsilon) \Gamma(1-\delta-2\epsilon)}.
\end{eqnarray}
There is a singularity for $k_\perp\to \infty$ for any $\nm k$. 
The pole at $\delta=0$ is an endpoint divergence from 
$\nm k\to 0$ for any $k_\perp$. This implies that $\np k$ becomes
large for fixed $k_\perp\sim \lambda^2$, and
hence the ``quark'' with momentum $k$ becomes collinear. 
In the soft region the transverse momentum and longitudinal
momentum integrals do not factorize, and there is also a divergence 
when $k_\perp\to \infty$ and $\nm k\to 0$ simultaneously, which
corresponds to the double pole in the hard-collinear integral.

Since we did
not regularize the hard-collinear contribution analytically, the
correct procedure is to expand first in $\delta$ and then in
$\epsilon$. In fact the pole
in $\delta$ cancels with the collinear contribution before expanding
in $\epsilon$. However, performing both expansions to compare 
with (\ref{cI}) we obtain
\begin{equation}
\label{sI2}
I_s =  -\frac{1}{2p'\cdot l} \left(
\left[\frac{1}{\delta}-\ln\frac{m^2}{\nu^2}\right]
\left[\frac{1}{\epsilon}-\ln\frac{m^2}{\mu^2}\right] 
-\frac{1}{\epsilon^2}+\frac{1}{\epsilon}\ln\frac{m^2}{\mu^2} -
\frac{1}{2}\ln^2\frac{m^2}{\mu^2}+\frac{5\pi^2}{12}\right)
\end{equation}
The expansion in $\delta$ has generated a double pole in $\epsilon$.

\paragraph{\it Adding up.} 
The singularity in $\delta$ cancels in the sum of the collinear and
soft integral
\begin{equation}
I_c+I_s =  -\frac{1}{2p'\cdot l} \left(
-\frac{1}{\epsilon^2}+\frac{1}{\epsilon}\ln\frac{2p'\cdot l}{\mu^2}
-\ln\frac{2p'\cdot l}{\mu^2}\ln\frac{m^2}{\mu^2}+
\frac{1}{2}\ln^2\frac{m^2}{\mu^2}+\frac{5\pi^2}{12}\right).
\end{equation}
Finally adding to this the hard-collinear contribution (\ref{hcI}), 
the singularity in $\epsilon$ also cancels, and 
we obtain 
\begin{equation}
I_c+I_s+I_{hc} =  -\frac{1}{2p'\cdot l} \left(
\frac{1}{2}\ln^2\frac{2p'\cdot l}{m^2}+
\frac{\pi^2}{3}\right),
\end{equation}
in agreement with the expansion (\ref{exactI}) of the full integral.
We conclude that in general hard-collinear, collinear
and soft momentum regions must be considered. In the scalar integral 
(\ref{defI}) all three regions contribute already to the leading term
in the expansion. Semi-hard modes with scaling 
$(\lambda,\lambda,\lambda)$ are not needed in this calculation, since
the corresponding integrals are scaleless.

In QCD the photon-vertex integral contains a numerator proportional 
to $\nm k$ which
suppresses the collinear region by a factor 
of $\lambda^2$ relative to the hard-collinear and soft region. 
For this reason it is sufficient to consider only
hard-collinear and soft configurations in the factorization theorem
for $B\to\gamma l\nu$ at leading power in $1/m_b$, as has been 
done in \cite{Descotes-Genon:2002mw,Lunghi:2002ju,Bosch:2003fc}. 
Hard-collinear modes are perturbative and can be integrated out, 
resulting in hard-scattering
kernels. Soft and collinear modes have virtuality $\lambda^4\sim
\Lambda^2$, and cannot be treated in perturbation theory. The
$1/m_b$ suppression of the collinear contribution in QCD implies 
that the hadronic structure of the photon is a sub-leading
effect in $B\to\gamma l\nu$ decay. 

\subsection{Off-shell regularization}

The scalar integral (\ref{cI}) has recently been discussed in 
\cite{Becher:2003qh}, however with $m=0$ and the external collinear and 
soft lines off-shell, $l^2\equiv -L^2 = \np l\nm l\sim \lambda^4$, 
and $(p')^2\equiv-(P')^2=\np p' \nm p'\sim \lambda^4$. It is instructive 
to discuss the difference to the case above. 

The hard-collinear integrals are identical as they should be, because the 
two integrals differ only at the small scale $\lambda^4$. The 
collinear contribution is now given by 
\begin{equation}
\label{cIpr}
I_c^\prime =  -\frac{1}{2p'\cdot l} \,\left(\frac{\nu^2}{2p'\cdot l}
\right)^\delta \int_0^1 \frac{du}{u^{1+\delta}}\,
\left(\mu^2 e^{\gamma_E}\right)^\epsilon
\int\frac{d^{d-2}k_\perp}{\pi^{d/2-1}}\,\frac{-1}{k_\perp^2-u(1-u)(P')^2}, 
\end{equation}
with $\np k=u\np p'$. This is to be 
compared to the first line of (\ref{cI}). When the 
transverse momentum integral is taken, it supplies a 
factor $[u (1-u)]^{-\epsilon}$, 
so the integral does appear to be defined in dimensional regularization 
($\delta=0$). However, at fixed transverse momentum there is still an 
endpoint divergence as $u\to 0$, which is not regularized dimensionally. 
Hence the integral has the same divergence as $k_\perp\to\infty$ and the 
same endpoint divergence as in the on-shell, massive case. In particular, 
the endpoint divergence cancels again with a soft endpoint divergence, at 
any transverse momentum. 

\begin{figure}[p]
   \vspace{-3.5cm}
   \hspace*{-0.4cm}
   \epsfysize=27cm
   \epsfxsize=19.5cm
   \centerline{\epsffile{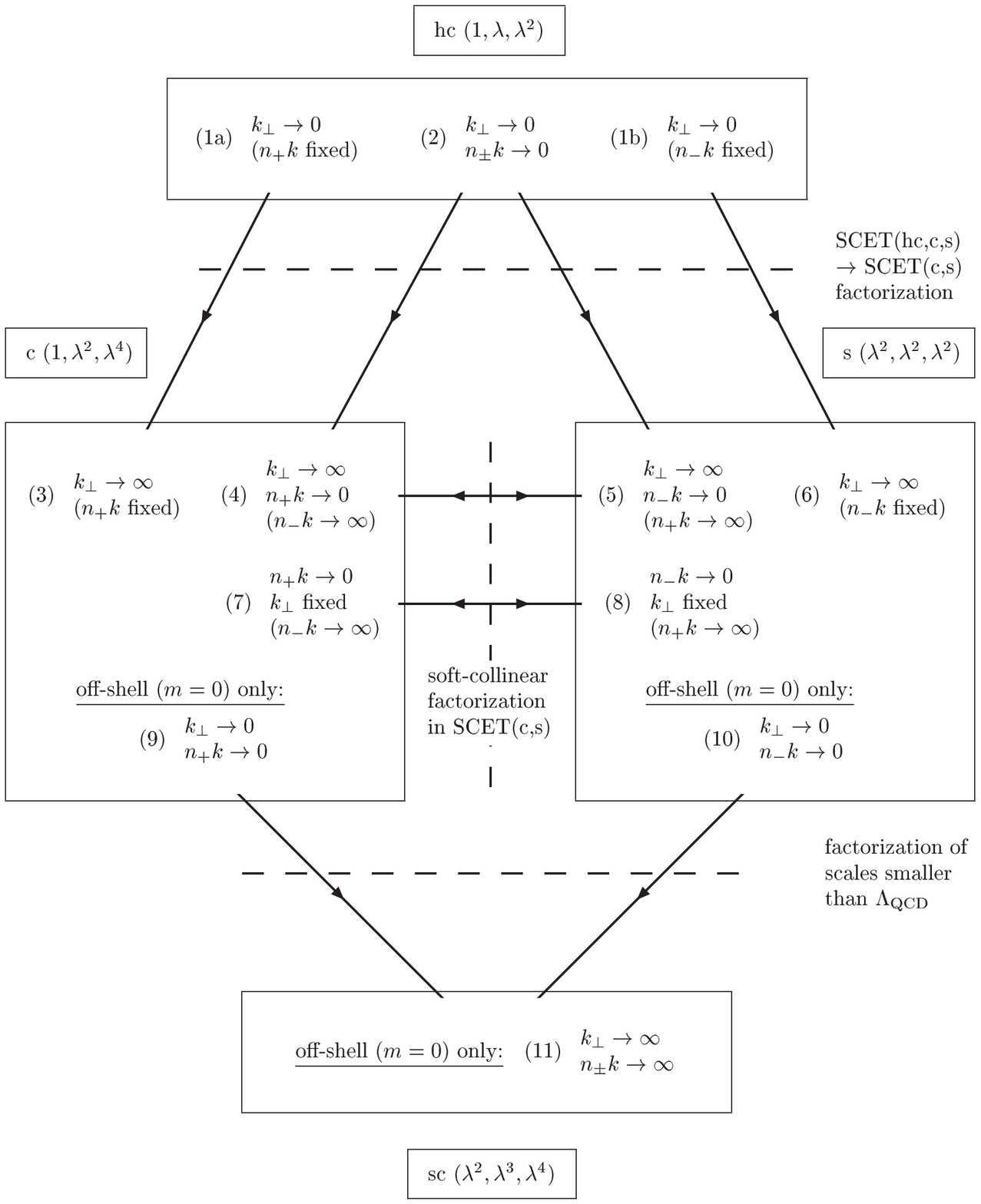}}
   \vskip-5cm
\centerline{\parbox{14cm}{\caption{\label{fig:divs}\small  
Divergence structure of the integral (\ref{defI}) and its 
off-shell, massless version when expanded by regions. The arrows 
indicate the divergences in different regions that are related 
and cancel each other. The dashed lines mark the various 
factorization steps.}}}
\end{figure}

The off-shell integral exhibits a new type of divergence, when 
$k_\perp\to 0$ and $\np k\to 0$ simultaneously. Since $k_\perp\to 0$ 
this must be related to a new momentum region with transverse momentum 
smaller than $\lambda^2$. A similar comparison of the soft integral 
shows that it possesses a singularity for $k_\perp\to 0$ and 
$\nm k\to 0$ in addition to those already discussed. In  
\cite{Becher:2003qh} it is shown that 
these extra divergences are compensated by a 
``soft-collinear'' contribution, where the loop momentum scales as 
$(\lambda^2,\lambda^3,\lambda^4)$, and that the sum of all four contributions 
reproduces the expansion of the full integral. The various divergences 
and their relations are summarized in Figure~\ref{fig:divs}.

Applied to effective theories of QCD, where $\lambda^2\sim \Lambda$, 
the various factorization steps shown in the Figure correspond 
to matching SCET onto an effective theory of only soft and 
collinear modes, and the factorization of soft and collinear modes 
within the low-energy theory. The third factorization step is needed only 
for the off-shell, massless case, and does not have an equivalent in 
QCD, since it concerns the separation of transverse momentum 
scales below $\Lambda$. 
Contrary to QCD, where we suppose that non-perturbative dynamics provides 
a universal infrared cut-off of order $\Lambda$, the off-shell 
regularization of a massless Feynman integral does not prevent  
sensitivity to arbitrarily small loop momenta.\footnote{It is very 
likely that with more loops, more modes of 
successively smaller virtuality must be introduced, 
with no lower limit on the virtuality as the 
number of loops increases.} This need not hinder the use of 
the off-shell regularization and the introduction of soft-collinear 
modes as a technical construct. This point of view has 
been taken in \cite{Becher:2003kh}. However, it is 
clear from (\ref{cIpr}) that the breakdown of naive 
soft-collinear factorization is related to endpoint divergences that 
already occur at transverse momenta of order $\lambda^2$, and is 
{\em per se} unrelated to the existence of soft-collinear modes. 
In dimensional regularization soft-collinear modes must be introduced 
as a way to recover the non-analytic dependence of the integral 
on $\np p' \nm l$. From a physical point of view, however, it is 
sufficient to describe the low-energy physics in terms 
of collinear and soft modes only. 

It is important to keep 
in mind that matrix elements in the low-energy effective theory 
cannot be naively factorized into soft and collinear matrix elements.
We may always factorize soft and collinear contributions 
in a perturbative Feynman integral with an appropriate regulator (as 
done above), but when $\lambda^2$ is set to $\Lambda$, one must examine 
whether factorization occurs at weak coupling. If it does not, 
factorization of soft and collinear modes cannot be implemented 
perturbatively in QCD. If endpoint divergences can be shown 
to be absent to all orders in perturbation theory, so that no 
regularization is required, this indicates that naive 
soft-collinear factorization is valid. 

\subsection{Interpretation of the result}

\paragraph{\it Hard-collinear modes in SCET.} 
The toy integral clarifies that SCET, defined as the effective 
theory after integrating out hard modes, contains two 
collinear modes with different virtuality. If SCET is formulated 
with a single collinear quark and gluon field, the collinear fields 
cannot be assigned an unambiguous scaling law, and power counting 
is no longer manifest in the vertices of the effective 
theory,\footnote{This also occurs in the standard formulation of 
non-relativistic effective theory, where the quark and gluon fields 
do not obey an unambiguous velocity scaling rule.} unless one of the 
two collinear modes is irrelevant for a specific process. An alternative  
is to introduce separate hard-collinear and collinear 
fields in SCET. The corresponding Lagrangian can be taken as 
a starting point for the second matching step, in which hard-collinear 
modes are integrated out and SCET is reduced to an effective Lagrangian 
for soft and collinear modes only. This formulation will be used 
in some of the technical steps in the following two sections. 

\begin{figure}[t]
   \vspace{-5.1cm}
   \epsfysize=25cm
   \epsfxsize=18cm
   \centerline{\epsffile{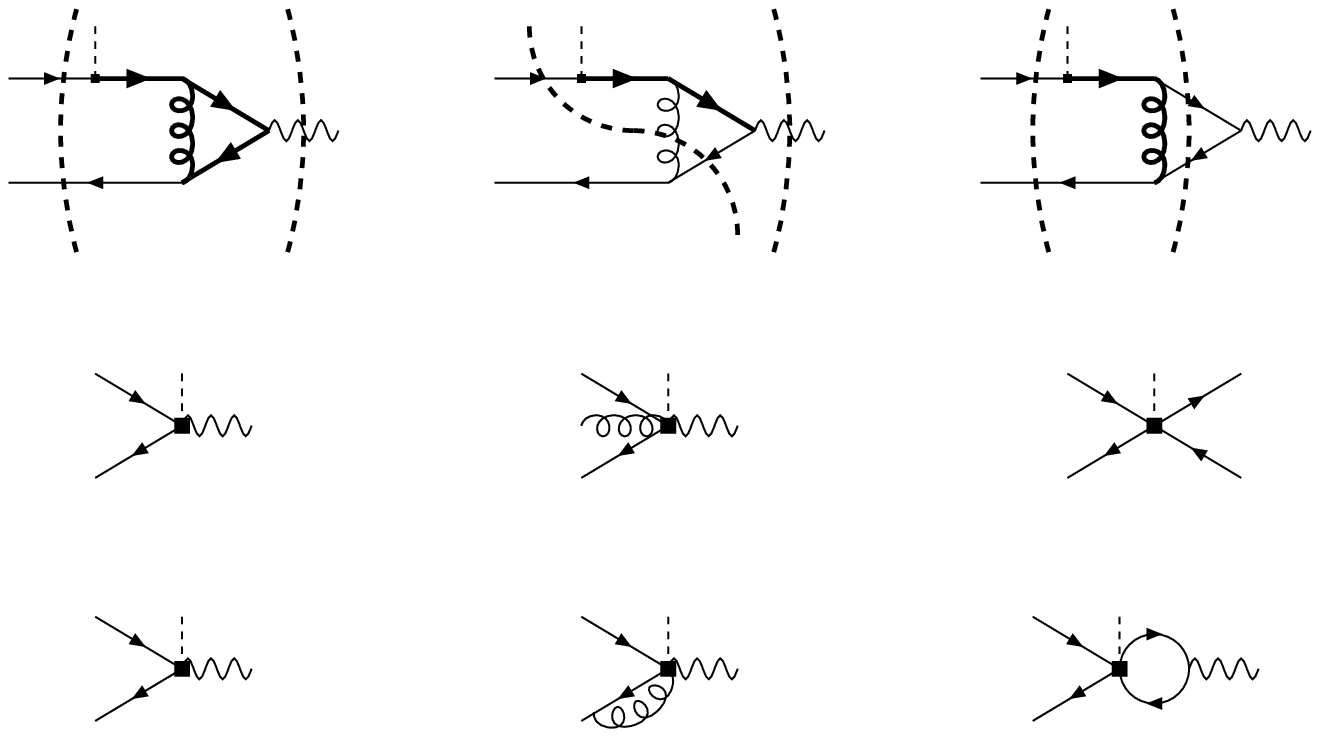}}
   \vspace*{-13cm}
\centerline{\parbox{14cm}{\caption{\label{fig:regions}\small  
Diagrammatic and operator/matrix element representation of the 
hard-collinear (left column), soft (middle column) and collinear 
contribution to the diagram of Figure~\ref{fig:vertex}. Each column 
shows: the original diagram with the hard-collinear subgraph 
marked by bold-face lines (upper row), and with the dashed line 
indicating where the graph factorizes into a short-distance and 
long-distance subgraph; the operator vertex in the effective theory 
corresponding to the contracted hard-collinear subgraph (middle row); 
the contribution to the operator matrix element 
$\langle \gamma|O_i|\bar q b\rangle$ corresponding to the original 
diagram (lower row).}}}
\end{figure}

A comment is necessary on the transverse momentum scaling of hard-collinear 
modes. When a soft and collinear momentum combine to a 
hard-collinear fluctuation of virtuality $\lambda^2$, the hard-collinear 
transverse momentum must be of order $\lambda^2$ by momentum conservation. 
This is the case for the hard-collinear propagators in 
the soft and collinear contributions to the toy integral (see the 
upper row of Figure~\ref{fig:regions}). However, in hard-collinear 
loops the transverse momentum is of order $\lambda$, as 
one can easily verify from the location of poles of the 
hard-collinear integrand. Assuming $k_\perp\sim \lambda^2$ 
would make all hard-collinear loop integrals vanish, since the integrals 
would have to be expanded in $k_\perp^2$. This would 
obviously fail to reproduce the expansion of the exact integral. 
We therefore assign the generic scaling $(1,\lambda,\lambda^2)$ to 
hard-collinear modes, as given in Table~\ref{tab1}. Non-generic scaling 
in tree subgraphs is not particular to the present case of 
hard-collinear modes. When one integrates 
out hard heavy quark fluctuations generated by the interaction of 
near on-shell heavy quarks with hard-collinear or collinear gluons 
\cite{Bauer:2000yr}, the off-shell modes have momentum 
$(1,\lambda,1)$ or $(1,\lambda^2,1)$, unlike the generic 
hard momentum $(1,1,1)$.

\paragraph{\it Operator interpretation of the toy integral.} 
We proceed to discuss the three contributions to the toy integral 
in terms of operators and matrix elements of an effective theory 
for soft and collinear modes. This discussion will be heuristic, since 
we abstract from the scalar integral and use QCD terminology, but 
without making the notation completely explicit. 

We imagine that Figure~\ref{fig:vertex} represents a correction to 
the matrix element $\langle\gamma| J | \bar q b\rangle$ of the 
$b\to u$ transition current between a $\bar q b$  state 
with fixed light quark momentum $\nm l$ and 
a photon with large energy $E=\np p'/2$. 
The corresponding tree diagram has one 
hard-collinear line joining the weak vertex to the photon vertex. In the 
effective theory (of soft and collinear modes) this is written as 
\begin{equation}
\label{c0tree}
C_0(E,\nm l) \,
{}_{\rm FT}\langle\gamma|[A_\gamma(s\np)]_c\,
[\bar q(t\nm) h_v(0)]_s | \bar q b\rangle(E,\nm l).
\end{equation}
The symbol ${}_{\rm FT}\langle \ldots\rangle$ means that a 
Fourier transform of the matrix element with respect to the position 
arguments of the fields is taken, with $(E,\nm l)$ the variables conjugate 
to $(s\np,t\nm)$. The index on products of fields indicates whether 
they are soft or collinear, and the non-locality of the operator 
is related to the non-polynomial dependence of 
the hard-collinear propagator on the momentum component $\nm l$ 
of the light external quark, and the momentum 
component $\np p'$ of the external photon. 
The matrix element factorizes trivially into  
\begin{equation}
{}_{\rm FT}\langle\gamma|[A_\gamma(s\np)]_c|0\rangle(E)\,
{}_{\rm FT}\langle 0|
[\bar q(t\nm) h_v(0)]_s | \bar q b\rangle(\nm l).
\end{equation}
The photon matrix element can be calculated. When the 
$\bar q b$ state is replaced by a $\bar B$ meson, the soft matrix element 
gives the $B$ meson light-cone distribution amplitude. Hence 
(\ref{c0tree}) assumes the form of a convolution of a hard-collinear 
coefficient function with the $B$ meson light-cone distribution function, 
which reproduces the factorization property 
of the $B\to \gamma$ transition at leading order in $1/m_b$, and 
at leading order in $\alpha_s$ \cite{Korchemsky:1999qb}.

The hard-collinear contribution to the toy integral and its operator 
interpretation is shown in the left column of Figure~\ref{fig:regions}. 
When the hard-collinear subgraph is contracted to a ``point'', the 
corresponding operator has the same field content as in (\ref{c0tree}), 
but with a different coefficient function 
$C_1(E,\nm l)$ as a result of the 
loop integration. We identify $C_1$ as a 1-loop correction to the 
hard-scattering kernel. The explicit calculation shows that there 
is a double pole in $1/\epsilon$, leaving a double-logarithmic dependence 
on the factorization scale $\mu$.

Consider now the soft contribution (middle column in the Figure). The 
hard-collinear subgraph has an additional external soft gluon line, so 
the operator in the effective theory has the structure 
$[\bar q A h_v]_s\,[A_\gamma]_c$ (second line in the Figure). The 
matrix element in the third line 
of Figure~\ref{fig:regions} takes the form 
\begin{equation}
\label{O2}
{}_{\rm FT}\langle\gamma|[A_\gamma(s\np)]_c|0\rangle(E)
\int d\omega\,C_2(E,\nm l,\omega)\,
{}_{\rm FT}\langle 0|[\bar q(t_2 \nm) A(t_1\nm) h_v(0)]_s|\bar q b
\rangle(\nm l,\omega).
\end{equation}
The soft matrix element can be identified 
with a three-particle light-cone distribution amplitude $\phi_{\bar q b g}$
of the $\bar q b$ state. Comparing this expression to (\ref{sI}), 
we see that the $\nm k$ integral in (\ref{sI}) 
corresponds to the integration 
over $\omega$, while the transverse momentum integral is included in 
the definition of the light-cone distribution amplitude. 

In the conventional hard-scattering formalism the scale dependence 
of the distribution amplitude would cancel against the scale dependence 
of a hard-scattering kernel (such as $C_1$). This cannot be completely 
correct here, since the $\omega$-integral has an endpoint divergence 
as $\omega\to 0$, which corresponds to the $1/\delta$ singularity 
in (\ref{sI}). The associated $\nu$-dependence is 
not cancelled by a hard-scattering kernel, but by the collinear 
contribution as seen from the toy example. The existence of an 
endpoint divergence implies that expression (\ref{O2}) in its entirety 
has a scale-dependence 
different from the two matrix elements in the factorized expression. 
This possibility is not considered in the conventional hard-scattering 
formalism.\footnote{In the leading power analysis of $B\to\gamma l\nu$ 
decay there is no endpoint divergence from the soft contribution 
of the photon vertex integral \cite{Descotes-Genon:2002mw}. 
The reason for this is that the corresponding operator is 
$[\bar q \nm A h_v]_s\,[A_\gamma]_c$, which is related to 
$[\bar q h_v]_s\,[A_\gamma]_c$ by gauge invariance. It is in fact included 
in the gauge-invariant definition of $[\bar q h_v]_s\,[A_\gamma]_c$, 
which contains a path-ordered exponential. In sub-leading power, 
the gluon can be transverse, and an endpoint divergence 
appears. This is consistent with the fact that the collinear 
contribution to the photon vertex is also sub-leading power in QCD.}

The operator interpretation of the collinear integral is illustrated in 
the third column of Figure~\ref{fig:regions}. The photon line is not 
directly connected to the hard-collinear subgraph in this case. 
Rather the operator that results after contracting the hard-collinear 
subgraph has field content 
$[\bar q h_v]_s\,[\bar q q]_c$ (second line in the Figure). 
The matrix element (third line) can be written as 
\begin{equation}
\label{O3}
{}_{\rm FT}\langle 0|[\bar q(t \nm)h_v(0)]_s|\bar q b\rangle(\nm l)
\int_0^1 du\,C_3(E,u,\nm l)\,
{}_{\rm FT}\langle\gamma|[\bar q(s_1\np)q(s_2\np)]_c|0\rangle(E,u).
\end{equation}
This seems to represent the convolution of a hard-scattering kernel $C_3$ 
with the two-particle light-cone distribution amplitude of 
the $\bar q b$ state  $\phi_{\bar q b}$ and the $q\bar q$ light-cone 
distribution amplitude of the photon $\phi^\gamma_{q\bar q}$. 
This is only correct with the understanding that the $u$-integral is  
divergent and must be regularized in a way that is consistent 
with the regularization of the $\omega$-integral in the soft contribution. 
The additional divergence, which is not related to the renormalization 
of the conventional light-cone distribution amplitudes, is 
the endpoint divergence of the $\np k$ integral in (\ref{cI}). 
The associated $\nu$-dependence cancels against the $\nu$-dependence 
of the soft contribution. In general, the distribution 
amplitudes may themselves depend on the additional regularization, 
and hence differ from the distribution amplitudes that 
appear in the hard-scattering formalism. 

To summarize this discussion, we 
distinguish two steps of factorization. In the first step, we 
integrate out the large-virtuality hard-collinear modes, and represent 
the result in terms of operators of soft and collinear fields. 
These operators will be non-local, reflecting the fact that 
the hard-scattering kernels appear in convolutions rather than as 
multiplicative factors. The second factorization step refers 
to the separation of soft and collinear modes  
{\em within} the effective theory of soft and collinear modes. 
In our example, the photon couples only to collinear lines, and the 
$\bar q b$ state couples only to soft lines, so we would expect 
the effective theory matrix elements to factorize into a 
matrix element of collinear fields between the photon and the 
vacuum, and a matrix element of soft fields between the vacuum 
and the $\bar q b$ state. If this were the case, the process would 
factorize into $S\star T\star \Phi$. The factorization scale dependence 
of the soft factor $S$ and of the collinear factor $\Phi$ 
would cancel separately with that of the hard-scattering kernel $T$, 
but the soft and collinear factors would be unrelated. 
The endpoint divergences prevent such a complete factorization. For 
our toy example we find instead a factorization formula that takes 
the schematic form
\begin{equation}
\langle \gamma| J| \bar q b\rangle = 
(C_0+C_1)\star \phi_{\bar q b} + 
\Big[C_2\star \phi_{\bar q b g}\Big]_\nu+
\Big[\phi^\gamma_{q \bar q}\star C_3\Big]_\nu\star 
\phi_{\bar q b}.
\end{equation}
The first term on the right-hand side represents a direct photon 
contribution; in the third term the partonic structure of the photon 
is resolved. The square brackets indicate the additional scale-dependence 
introduced by the endpoint divergences, which connect the second with the  
third term. If the scale $\nu$ is chosen such that the third term 
contains no large logarithm related to the endpoint divergence, 
we can interpret it as a endpoint-subtracted hard-scattering 
contribution to the $\bar q b\to\gamma$ transition. For our toy 
integral, (\ref{cI},\ref{sI2}) show that this corresponds to taking 
$\nu^2$ of order $2 p'\cdot l$. The corresponding endpoint  
logarithm then resides in the second term, which we may call the  
``soft overlap'' contribution (since a soft line connects the 
initial state with the photon as seen from the middle column of 
Figure~\ref{fig:regions}). The two terms are related via 
their $\nu$-dependence, such that the sum is independent of 
the implementation of soft-collinear or ``endpoint'' factorization. 
A similar structure is expected for the $B\to\pi$ form factor 
\cite{Beneke:2000wa}.


\section{\boldmath Heavy-to-light transitions in $\SCETII$}
\label{sec:tree}

\begin{figure}[t]
   \vspace{0.5cm}
   \epsfxsize=10cm
   \centerline{\epsffile{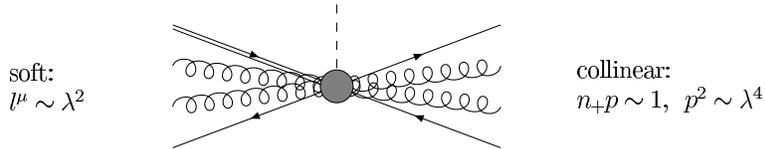}}
   \vspace*{0.3cm}
\centerline{\parbox{14cm}{\caption{\label{fig:setup} 
Kinematics of an exclusive heavy-to-light transition in 
$\SCETII$. The heavy quark and the soft partons in the 
$B$ meson must be converted into a cluster of collinear 
partons.}}}
\end{figure}

The effective theory representation of the heavy-to-light transition 
currents $\bar \psi\hspace{0.04cm}\Gamma Q$ 
is obtained in two steps: first the 
hard modes are integrated out, and the current is described 
in soft-collinear effective theory including hard-collinear modes. 
We shall denote this theory by $\SCETI$ (also called SCET${}_{\rm I}$ in 
the literature \cite{Bauer:2002aj}). This step, 
in which it is not necessary to distinguish hard-collinear and
collinear, has already been 
discussed in \cite{Bauer:2000yr,BCDF}.  
We will be mainly concerned with the second 
matching step, in which the hard-collinear modes are integrated out  
and the transition current is finally represented in terms of operators
constructed only from soft and collinear fields. We refer to the 
theory of soft and collinear fields as $\SCETII$ (also called 
SCET${}_{\rm II}$). The kinematics of a  heavy-to-light transition is
illustrated in Figure~\ref{fig:setup}. In contrast to \cite{BCDF} the 
invariant mass of the final state is restricted to order
$\lambda^4\sim \Lambda^2$ as appropriate to an exclusive decay. This
implies that the final state must now consist only of collinear lines, 
since the addition of a soft line would increase the virtuality to
$\lambda^2$. The initial state is described by a heavy quark and 
soft lines with total invariant mass near $m_b^2$. The $\SCETII$ transition
current has to turn a cluster of soft modes with the quantum numbers
of the $\bar B$ meson into a cluster of collinear lines with the
quantum numbers of the final-state meson.  

We begin with a brief description of the $\SCETII$ fields, gauge
symmetries and Lagrangian. This theory is in many ways simpler than 
$\SCETI$, because the dynamics of the soft-collinear transition
resides only in the effective current. We then discuss in detail the
representation of the heavy-to-light current 
for an exclusive decay. In this section we restrict ourselves to
tree-level matching. The general case will be considered in 
Section~\ref{sec:formfactor} to the extent that is necessary to prove
the factorization of form factors. However, we briefly sketch the
structure of transition operators and their coefficient functions 
beyond tree level at the end of this section. 

\subsection{\boldmath Elements of $\SCETII$}

\paragraph{\it Fields.} 
The $\SCETII$ Lagrangian and operators are built from a collinear 
light quark field $\xi_c$, a collinear gluon field $A_c$, and 
soft light quark, heavy quark and gluon fields, denoted by 
$q_s$, $h_v$, $A_s$, respectively. As in \cite{BCDF} we assume that 
collinear fields describe particles with large momentum in the 
direction of the light-like vector $n_-$. $n_+$ is another light-like
vector, satisfying $\nm \np=2$, and $v$ will be the velocity vector 
labelling soft heavy quark fields. We will present our results in
a general frame subject to the conditions 
$n_+ v \sim 1$, $n_- v \sim 1$, $v_\perp \sim \lambda^2$.

The scaling of quark and gluon fields can be read off
from the corresponding propagators in momentum space. 
For the quark fields one finds 
\begin{equation}
\xi_c = \frac{\slash n_- \slash n_+}{4} \, \psi_c\sim \lambda^2, \qquad
h_v =  \frac{1+\slash v}{2} \, Q_v \sim \lambda^3, \qquad
q_s \sim \lambda^3. 
\end{equation}
The ``opposite'' projections 
of the full collinear quark field $\psi_c$ and the full soft heavy
quark fields $Q_v$ (defined as the heavy quark field with the rapid 
variations $e^{-i m_b vx}$ removed) 
are $\lambda^2$ suppressed, and integrated out. 
Gluon fields scale as the corresponding derivatives, 
\begin{equation}
  n_+ A_c \sim 1, \qquad
  A_{\perp c} \sim \lambda^2, \qquad
  n_- A_c \sim \lambda^4, \qquad
  A_s \sim \lambda^2.
\end{equation}
In deriving this, we used that the integration measure 
$d^4x \sim 1/\lambda^8$, when the integral is over products 
of only collinear fields or products of only soft fields. 
This follows from the fact that an $x$-component scales inversely to
the corresponding momentum component.

\paragraph{\it Multipole expansion.} 
Since soft and collinear fields have significant variations over
different length scales in the
$n_-$ and $n_+$ directions, 
they have to be multipole-expanded in products of soft and 
collinear fields. The multipole expansion
in $\SCETII$ is different from the multipole expansion defined 
in \cite{BCDF}, which applies to a theory with hard-collinear and 
soft fields, and no collinear fields. Here we need 
\beq
 \phi_s(x) &=& 
      \phi_s(x_-) 
      + \frac{n_- x}{2}
        \left[ n_+\partial \, \phi_s \right](x_-)
      + \ldots,
\nonumber \\[0.2em]
 \phi_c(x) &=& 
      \phi_c(x_+) 
      + \frac{n_+ x}{2} 
        \left[ n_-\partial \, \phi_c \right](x_+)
      + \ldots,
\label{eq:multipole}
\eeq
where 
\beq
&& x_- \ \equiv \  n_+ x \, \frac{n_-}{2} + x_\perp,
\qquad x_+ \ \equiv \ n_- x \, \frac{n_+}{2} + x_\perp.
\label{xpm}
\eeq
The correction terms in the two expansions of 
(\ref{eq:multipole}) are both $\lambda^2$ suppressed relative to the
leading terms. 

\paragraph{\it Gauge symmetry.}
The effective theory should be invariant under collinear and 
soft gauge transformations, defined as the restriction of 
gauge functions $U(x)$ to the corresponding spatial variations. 
The implementation of gauge
transformations in the effective theory is not unique, 
since field redefinitions or applications of the
field equations can be used to alter the gauge-transformation 
properties \cite{BCDF,Beneke:2002ni,Chay:2002mw}. 

In $\SCETII$ collinear and soft fields decouple at leading 
power in the $\lambda$ expansion as will be seen below. 
Furthermore, the product of a 
collinear and a soft field has hard-collinear momentum modes, 
therefore general soft gauge transformations acting on collinear fields 
(and vice versa) are not allowed. A natural choice is then 
to define \cite{Hill:2002vw}
\beq
&& \xi_c  \ \to \ U_c \, \xi_c, \qquad   
A_c \ \to \ U_c A_c U_c^\dagger 
       + \frac{i}{g} U_c \left[\partial,U_c^\dagger\right], 
\nonumber \\[0.3em]
&& h_v \ \to \ U_s \, h_v, \qquad
q_s  \ \to \ U_s \, q_s, \qquad \  
A_s \ \to \ U_s A_s U_s^\dagger 
       + \frac{i}{g} U_s \left[\partial ,U_{\rm
 s}^\dagger\right].
\label{eq:gauge-soft}
\eeq

\paragraph{\it Lagrangian.} The $\SCETII$ Lagrangian describes 
interactions of soft and collinear fields. In general, there can be
scattering processes of the type $s+c\to s+c$. It has been shown 
in \cite{Hill:2002vw} for quark scattering that these interactions 
are power-suppressed. Below we extend this to gluons and derive the
explicit form of the leading power-suppressed interactions. Because 
of the decoupling of soft and collinear modes at leading power, 
the Lagrangian is simply 
\begin{equation}
{\cal L}^{(0)} =  {\cal L}_s+{\cal L}_c 
\end{equation}
at leading power, with 
\begin{eqnarray}
{\cal L}_{c} &=& - \frac{1}{2} \, {\rm tr} 
\left(F_{\mu\nu c} F^{\mu\nu}_c\right)+ 
\bar\xi_c \left( i n_- D_c + (i \Slash D_{\perp c} -m) \, 
\frac{1}{in_+ D_c} \, (i \Slash D_{\perp c} +m)
\right) \frac{\slash n_+}{2} \, \xi_c,
\nonumber\\
{\cal L}_{s} &=& - \frac{1}{2} \, {\rm tr} 
\left(F_{\mu\nu s} F^{\mu\nu}_s\right)+
 \bar q_s \, (i \Slash D_s - m) \, q_s +{\cal L}_{\rm HQET}
\label{leadinglag}
\end{eqnarray}
and $m\sim \lambda^2$ the light quark mass.
The heavy quark interactions with soft fields are described by the standard 
heavy quark effective theory (HQET) Lagrangian ${\cal L}_{\rm HQET} = 
\bar h_v\,i v\cdot D_s \,h_v+\ldots$ 

For the kinematic situation shown in Figure~\ref{fig:setup} 
scattering processes $s+c\to s+c$ cannot occur. On the other hand, 
the crossed process $s+s\to c+c$ is not possible by momentum 
conservation in the $n_\mp$ directions, since $\np p>0$ for a
collinear momentum and $\nm l>0$ for a soft momentum.  
It follows that insertions of the soft-collinear
interaction terms from the sub-leading Lagrangian have zero 
$\langle \pi|\ldots|\bar B\rangle$ matrix elements, so that we can
simply work with the Lagrangian without soft-collinear interactions to any 
accuracy. (See \cite{Becher:2003qh} for a related discussion.) 
Beyond leading power there are additional terms 
in the collinear Lagrangian ${\cal L}_c$,  
which arise upon integrating out heavy-quark
loops with external collinear lines. Heavy-quark loops also 
generate additional soft interaction terms, which 
correct ${\cal L}_s$, which also includes the $1/m_b$ suppressed 
terms from the HQET Lagrangian. However, the
soft-collinear interactions all reside in the effective current.

\paragraph{\it States and hadronic matrix elements.}

The $B$ meson states in the effective theory may be defined as the
eigenstates of the leading-order soft Hamiltonian. These states are
identical to those used in HQET. Alternatively, if we regard the
states as the eigenstates of the exact soft Hamiltonian, these states
coincide with the $B$ meson states of full QCD, since the soft 
Lagrangian to all orders is equivalent to the full QCD Lagrangian. 
Similarly, the light meson state in $\SCETII$ may be defined as 
the eigenstate of the leading-order collinear Hamiltonian. This
Hamiltonian is equivalent to QCD without heavy quarks, so the 
pion state in the effective theory is the same as in QCD (without
heavy quarks). Defining the pion with respect to the exact collinear 
Hamiltonian implies that the pion state is the same as in full QCD
with heavy quarks. In the following we adopt the convention that the
states are defined with respect to the exact soft or collinear
Hamiltonians, so that we do not distinguish the effective theory 
states from those in QCD. It would be a simple matter to make 
the $\lambda$ dependence of the states explicit. 
An implicit
assumption here is that the separation of collinear and soft modes is
done without an explicit cut-off (dimensional or analytic 
regularization). With a regularization that breaks boost invariance
the collinear Lagrangian is not equivalent to the full 
QCD Lagrangian. 
From the conventional normalization of hadronic states it 
follows that
\beq
&&  |B(p)\rangle \sim \lambda^{-3}, 
\qquad |\pi(p')\rangle\sim \lambda^{-2},
\label{eq:hadron-scaling}
\eeq
where the light meson is assumed to be energetic. 

That a pion is made 
only of collinear partons can be understood by 
noting  that a collinear pion state 
can be obtained from a pion at rest by a large Lorentz boost. 
In a pion at rest all partons have momenta of order $\Lambda$, 
so the boosted system contains only collinear modes. 
Adding a soft parton to the collinear modes produces a configuration 
of invariant mass $m_b \Lambda $, which cannot contribute to the 
pion pole. On the other hand, with the same line of reasoning,
a $B$ meson consists only of soft partons (and a heavy quark), 
since adding a collinear 
mode produces a configuration far away from the $B$ meson pole. 

An apparent consequence of the 
absence of soft-collinear interactions and the nature of the states 
is that an expression 
\begin{equation}
C*\langle \pi|f(\phi_c)g(\phi_s)|\bar B\rangle
\end{equation}
where $f(\phi_c)$ ($g(\phi_s)$) is a non-local product of collinear (soft) 
fields and the star denotes convolutions, factorizes into 
\begin{equation}
\langle \pi|f(\phi_c)|0\rangle *C* \langle 0|g(\phi_s)|\bar B\rangle.
\end{equation}
As shown in Section~\ref{sec:toy} this should be considered as formal,
since the collinear and soft convolution integrals can be divergent.

Another consequence is that the QCD current  matrix element 
$\langle \pi(p')|\bar u \, \Gamma \, b|B(p)\rangle$ simply matches 
to  $\langle \pi(p')|J_{\rm eff}|B(p)\rangle$, so the problem
reduces to obtaining the effective current. Already at this point we 
may note that the apparently leading term vanishes, 
\begin{equation}
\langle \pi(p')|\bar \xi_c \, \Gamma \, h_v|\bar B(p)\rangle = 0, 
\label{eq:example}
\end{equation}
because the quantum numbers of the product of collinear fields (here 
the single field $\bar \xi_c$) do not match those of a pion, and 
the quantum numbers of the soft fields (here only $h_v$) do 
not match those of the $\bar B$ meson. This can be formalized by 
saying that the effective Lagrangian is invariant under separate phase 
transformations of the collinear, soft and heavy quark fields, 
so we can assign ``collinear quark number''
to products of operators, with $\xi$ fields carrying  
collinear quark charge $+1$, $\bar\xi$ having charge $-1$ and 
all other fundamental fields charge $0$. We shall see later that 
the first non-zero matrix element is 
suppressed by three powers of $\lambda$. This is the origin 
of the well-known $1/m_b^{3/2}$ suppression of heavy-to-light 
form factors at large recoil \cite{Chernyak:ag}. 

This leads to the important observation \cite{Bauer:2002aj} 
that power-suppressed currents in the effective theory become 
relevant to the $B\to\pi$ form factor at leading power. 
The derivation of these currents in $\SCETII$ will be worked out 
at tree level below, and in more generality but less explicitly in 
Section~\ref{sec:formfactor}. Note that in the intermediate $\SCETI$,  
where only hard modes are
integrated out, the power-counting for hadronic
matrix elements is not explicit. In particular in \cite{BCDF} 
the pion state included
soft-collinear interactions and the suppression of the matrix 
element of $\bar \xi_{hc}\,\Gamma\,
h_v$ (with $\xi_{\rm hc}$ denoting the hard-collinear quark field)   
with these pion states was not determined (see the
discussion at the end of Section 5.3 of \cite{BCDF} and 
in \cite{Hill:2002vw}). 

\paragraph{\it Light-cone gauge and Wilson lines.}

It will sometimes be convenient -- especially for the following 
tree-level matching of $\SCETI$ to $\SCETII$ --  to choose the
gauge
\begin{equation}
n_+ A_{hc} = n_+ A_c = n_- A_s = 0.
\label{eq:fixed-gauge}
\end{equation}
The usefulness of $n_+ A_{hc}=n_+ A_c = 0$ gauge follows from the fact that 
$\SCETI$ is non-local only due to the presence of Wilson lines 
in the direction of $n_+$, and due to the appearance of 
$(i\np\partial)^{-1}$. With 
$n_+ A_{hc} = \np A_{c}=0$ all Wilson lines reduce to 1, 
and there are no fields of order 1. The usefulness of 
$\nm A_s=0$ gauge is related to the fact that 
in $\SCETI$ soft gluons decouple from 
collinear and hard-collinear gluons at leading order in $\lambda$ in 
this gauge. 

Once a particular result has been derived in this gauge, the 
collinear and soft gauge invariance is recovered by transforming the 
fields back to a general gauge using the gauge transformations
(\ref{eq:gauge-soft}).
As explained in \cite{Beneke:2002ni}, the transformation
matrices $U_c$ and $U_s$ that accomplish this are the 
light-like Wilson lines
\beq
&& U_c^\dagger(x) = W_c(x) = P\, 
  \exp\left(ig \int_{-\infty}^0ds \, n_+ A_c(x+ s n_+)\right),
\nonumber \\[0.2em]
&&  U_s(x) =   Y_s^\dagger(x) = P \, 
  \exp\left(ig \int^{\infty}_0dt \, n_- A_s(x+ t n_-)\right), 
\eeq
and the corresponding gauge transformation of the fields can be written as
\beq
&& \xi_c  \ \to \ W^\dagger_c  \xi_c, \qquad   
g A_c \ \to \ W_c^\dagger \left[i D_c W_c\right] \ \equiv \ {\cal A}_c,
\nonumber \\[0.3em]
&&
h_v \to Y_s^\dagger  h_v, \qquad
q_s  \ \to \ Y_s^\dagger  q_s, \qquad \  
gA_s \ \to \ Y_s^\dagger\left[ i D_s Y_s\right] \ \equiv \ {\cal A}_s.
\label{eq:cal-soft}
\eeq 
(Here and in the following derivatives in square brackets act only on 
the expression to their right inside the bracket.)
Because the Wilson lines transform as
\beq
&&  
Y_s \stackrel{U_c}{\to} Y_s, \qquad 
Y_s \stackrel{U_s}{\to} U_s Y_s, \qquad 
W_c \stackrel{U_c}{\to} U_c W_c, \qquad 
W_c \stackrel{U_s}{\to} W_c,
\eeq
the expressions on the right-hand side of
(\ref{eq:cal-soft}) are gauge-singlets. The fields ${\cal A}_{c}$, 
${\cal A}_{s}$ 
have been introduced in \cite{Hill:2002vw} as building
blocks for manifestly gauge-invariant operators. 

\begin{figure}[t]
   \vspace{0.4cm}
   \epsfxsize=10cm
   \centerline{\epsffile{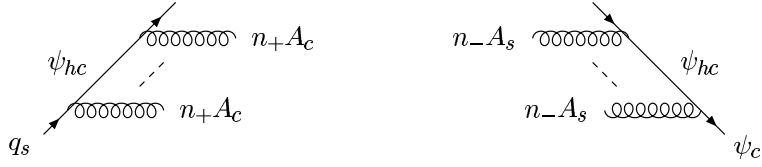}}
   \vspace*{0.3cm}
\centerline{\parbox{14cm}{\caption{\label{fig:Wilson}\small
Infinite sets of Feynman graphs with attachments of soft
gluons to collinear quarks and vice-versa. Integrating out
the intermediate hard-collinear propagators leads to the
Wilson lines $Y_s^\dagger$ 
and $W_c$, respectively, see (\ref{eq:Wilson}).}}}
\end{figure}

In a general gauge the Wilson lines emerge automatically 
from matching an infinite set of unsuppressed tree-level Feynman
diagrams with attachments of $n_+ A_c$ to soft fields, 
and $n_- A_s$ gluons to collinear fields as sketched in 
Figure~\ref{fig:Wilson}. Indeed, at leading power these diagrams 
result in\footnote{We do not write out the 
$+i\epsilon$ prescription on the propagators, which reads 
$+i\epsilon$ for $1/(i\np\partial)$ and $-i\epsilon$ for 
$1/(i\nm\partial)$. This follows, because the internal hard-collinear 
propagators are always space-like, with 
$(p'-l)^2\simeq -\nm l\,\np p'<0$, where $\np p'>0$ describes an 
outgoing collinear momentum, and $\nm l>0$ an incoming soft 
momentum. Hence in position space 
$1/(i\np\partial \,i\nm\partial+i\epsilon)$ 
is $1/(i\np\partial+i\epsilon)\, 1/(i\nm\partial-i\epsilon)$.}
\beq
&& \bar \psi_c \left(1 - g\Slash A_s\,\frac{1}{i\Slash D-m} \right)
  \simeq
    \bar \xi_c \left(1 + 
gn_- A_s \,\frac{1}{i n_- \overleftarrow{D}_s} \right)
  \simeq  \bar \xi_c\,Y_s^\dagger , 
\nonumber \\[0.2em]
&& \left(1 - \frac{1}{i\Slash D-m} \, g\Slash A_c\right)
  q_s \simeq
    \left(1 - \frac{1}{i n_+ D_c} \,gn_+ A_c \right)
  q_s \simeq W_c \, q_s. 
\label{eq:Wilson}
\eeq 
In practice using the fixed gauge (\ref{eq:fixed-gauge}) 
in intermediate steps and restoring the gauge symmetry 
via (\ref{eq:cal-soft}) is more efficient,
in particular when keeping track of Wilson lines arising
from multi-gluon vertices in the non-Abelian theory.

\subsection{Effective current at tree level}

We now consider the representation of heavy-to-light transition
currents $J=\bar\psi \hspace{0.04cm}\Gamma Q$, 
where $\Gamma$ is a Dirac matrix,
in $\SCETII$. We are only concerned with tree diagrams in this 
subsection. 

In the following we introduce different fields for hard-collinear and
collinear modes. We work with the gauge $n_+ A_c=\np A_{hc}=\nm A_s=0$ and 
present the gauge-invariant result only at the end. Integrating out 
hard intermediate heavy-quark propagators in tree diagrams,  
we obtain the current in $\SCETI$ in the form 
\begin{equation}
\label{jj}
J(x)=e^{-im_b v x}\,\left[\bar\psi\hspace{0.04cm}\Gamma{\cal Q}\right](x),
\end{equation}
where 
\begin{eqnarray}
\psi&=& \xi_c+\eta_c+\xi_{hc}+\eta_{hc}+q_s 
\nonumber\\
&=& \xi_c+\xi_{hc}+q_s-\frac{1}{i\np D_s}\,\frac{\slash n_+}{2} \, 
\left((i\Slash D_{\perp}+m)\,(\xi_c+\xi_{hc})+ 
(g\Slash A_{\perp c}+g\Slash A_{\perp hc}) \,q_s\right),
\nonumber \\[0.2cm] 
{\cal Q} &=& \left(1+\frac{i\Slash D_s}{2 m_b}\right) h_v - 
\frac{1}{\nm v}\frac{\slash n_-}{2 m_b}\,\left(
g\Slash A_{\perp c}+g\Slash A_{\perp hc}\right) h_v +
O(\lambda^4 h_v)
\label{Qfield}
\end{eqnarray}
with $\eta_{c}$, $\eta_{hc}$ the small components of the
(hard-)collinear spinor field. 
This corresponds to the expressions 
given in \cite{BCDF,Beneke:2002ni},
apart from the fact that we have made explicit the presence of
two different quark and gluon modes by substituting 
$\phi_c\to\phi_c+\phi_{hc}$ ($\phi=\xi,A$), and did not perform the multipole
expansion. 

Integrating out the hard-collinear fields at tree level expresses them
in terms of soft and collinear fields as solutions to the classical 
field equations. In the following we integrate out the hard-collinear 
quark field first and obtain the effective field $\psi$ depending of 
soft and collinear 
fields, and the hard-collinear gluon field. Inserting this into the gluon 
field equation determines the hard-collinear gluon field in terms of 
soft and collinear fields. Substituting back into the expression for 
$\psi$ also determines the quark field. 
The $\SCETII$ representation of the current 
at tree level then follows by inserting the expressions for the
hard-collinear fields into (\ref{Qfield}). We obtain the
hard-collinear fields below, but anticipate here that $\xi_{hc}\sim
\lambda^3$ and $A_{\perp hc}\sim \lambda^3$ for the classical fields. 
We used this to truncate the expansion of ${\cal Q}$ at order 
$\lambda^6$. None of the terms counted as $\lambda^2$ in \cite{BCDF} 
contribute to tree-level matching owing 
to the suppression of the classical hard-collinear fields.  Also 
when $\bar\psi$ and ${\cal Q}$ are multiplied together only those
terms with collinear quark number zero have non-zero 
$\langle\pi|\ldots|\bar B\rangle$ matrix elements. 

\subsubsection{Hard-collinear quark field}

We first integrate out the hard-collinear quark field, which can be
done exactly since the Lagrangian is bilinear in the quark field. This 
expresses the quark Lagrangian in terms of the soft and collinear 
fields and the hard-collinear gluon field. The latter is a function of
the soft and collinear fields to be determined later from the 
gluon field equation. 

The solution for the hard-collinear quark field can be derived in two 
ways, either by first integrating out the small 
component $\eta=\eta_c+\eta_{hc}$ of a collinear QCD spinor to 
go to $\SCETI$ and then integrating out $\xi_{hc}$ in  
$\SCETI$, or by integrating out the hard-collinear field 
$\psi_{hc}=\xi_{hc}+\eta_{hc}$ in QCD and then integrating out 
$\eta_c$. We briefly comment on the two procedures. 

In the first derivation, which is closer to the spirit of effective 
field theory, after eliminating $\eta$, we obtain the exact 
SCET Lagrangian before multipole expansion given as 
Eq. (6) of \cite{Beneke:2002ni}. In collinear 
light-cone gauge $\np A_c=0$, the quark Lagrangian reads 
\begin{eqnarray}
\label{collfinal}
{\cal L} &=&  
\bar\xi\, \left(i n_- D + [i \Slash D_\perp -m]\frac{1}{i n_+ D_s}\,
         [i\Slash D_\perp+m] \right)  \frac{\slash n_+}{2} \xi 
+ \bar{q}\, (i \Slash{D}_{s}-m)\, q  
\nonumber\\[0.2cm]
&& -\,\bar q_s \,g\Slash A_{\perp c}\frac{\slash
  n_+}{2}\, \frac{1}{in_+ D_s}\,g \Slash A_{\perp c} q_s
+ \bar \xi \,g\Slash A_{\perp c} q_s 
+ \bar \xi\,\frac{\slash n_+}{2} g\nm A_c q_s 
+ \bar\xi \,\frac{\slash n_+}{2}\,i \Slash D_\perp\, \frac{1}{in_+ D_s}\,
g\Slash A_{\perp c} q_s 
\nonumber\\[0.2cm]
&& +\,\mbox{h.c. of the $\bar\xi[\ldots]q_s$ terms.}  
\end{eqnarray}
Then we substitute $\xi\to\xi_c+\xi_{hc}$ and 
$A_c\to A_c+A_{hc}$ in (\ref{collfinal}) and integrate out 
$\xi_{hc}$. Not all of the interactions vertices generated by this 
substitution can be realized due to the requirement of momentum 
conservation. For example, 
the term $\bar \xi \Slash A_{\perp c} q_s$ from (\ref{collfinal}) 
is replaced by
\begin{equation}
(\bar \xi_c+\bar\xi_{hc})\,(\Slash A_{\perp c} +\Slash A_{\perp hc}) q_s 
\simeq 
\bar \xi_c\,\Slash A_{\perp hc} q_s+ 
\bar\xi_{hc}\,(\Slash A_{\perp c} +\Slash A_{\perp hc}) q_s. 
\end{equation}
Although vertices with a single hard-collinear field cannot contribute to 
hard-collinear loops by momentum conservation, they cannot be 
simply omitted from the effective Lagrangian, because soft and collinear 
momenta add up to a hard-collinear momentum with non-generic transverse 
momentum of order $\lambda^2$. In fact, integrating out a hard-collinear 
field $\phi_{hc}$ at tree-level implies a solution 
$\phi_{hc}=f(\phi_c,\phi_s)$ of the classical field equations that can 
have only fluctuations with transverse momenta of 
order of the transverse fluctuations of soft and collinear fields.  

In the second derivation directly from QCD we use the decomposition 
\beq
(i \Slash D-m)^{-1} &=& 
\frac{1}{in_+ D + (i \Slash D_\perp+m) \, \frac{1}{in_-D} \, 
(i \Slash D_\perp-m)}
\left( \frac{\slash n_+}{2} + (i \Slash D_\perp+m) \, \frac{1}{in_-D} \, 
\frac{\slash n_+ \slash n_-}{4} \right) 
\nonumber \\[0.2cm]
&&\hspace*{-1.5cm}+\, 
\frac{1}{in_- D + (i \Slash D_\perp+m) \, \frac{1}{in_+D} \, 
(i \Slash D_\perp-m)}
\left( \frac{\slash n_-}{2} + (i \Slash D_\perp+m) \, \frac{1}{in_+D} \, 
\frac{\slash n_- \slash n_+}{4} \right)
\label{Dinverse}
\eeq
of the inverse Dirac operator. We checked that the result for the 
hard-collinear quark field is the same in both methods.  
In both cases we found it useful to perform the calculation in 
light-cone gauge (\ref{eq:fixed-gauge}).

After eliminating the hard-collinear quark field as described, 
the result is expanded in powers of $\lambda$. 
In presenting the result of expansion we anticipate 
that the tree-level hard-collinear gluon field scales as 
$A_{\perp hc}\sim \lambda^3$ and $\nm A_{hc}\sim \lambda^4$ plus 
sub-leading terms. Below $A_{hc}^{(n)}$ denotes the term of 
order $\lambda^n$ in the solution for the hard-collinear gluon field. 
Since the effective current requires the expression for $\psi$ 
(see (\ref{Qfield})), we present directly the result for  
\begin{equation}
\label{psifield}
\psi = \xi_c+\eta_c+\xi_{hc}+\eta_{hc}+q_s \equiv 
\psi^{(2)} + \psi^{(3)} + \psi^{(4)} + \psi^{(5)} + \ldots
\end{equation}
up to order $\lambda^5$, which we need later. The expansion 
begins with $\psi^{(2)}=\xi_c$, and the sub-leading terms read
\begin{eqnarray}
\psi^{(3)} &=& \left(1+\frac{1}{in_-\partial}\,
  g\Slash A_{\perp c}\,\frac{\slash n_-}{2}\right)q_s,
\nonumber\\[0.2cm]
\psi^{(4)} &=& 
 \frac{1}{in_+\partial}\,
 (i\Slash D_{\perp c}+g\Slash A_{\perp s}+m)\,\frac{\slash n_+}{2}\,\xi_c 
\nonumber\\[0.2cm]
  &&-\,\frac{1}{in_-\partial}\,\left((i\Slash D_{\perp c}+
  g\Slash A_{\perp s}+m) 
  \,\frac{1}{in_+\partial}\,g\Slash A_{\perp s}+g\Slash A_{\perp s}\,
  \frac{1}{in_+\partial}\,(i\Slash D_{\perp c}-m)\right)\xi_c
\nonumber\\[0.2cm]
  &&-\,\frac{1}{in_-\partial}\,g\nm A_{hc}^{(4)}\,\xi_c 
  +\,\frac{1}{in_-\partial}\,g\Slash A_{\perp hc}^{(3)}\,
  \frac{\slash n_-}{2}\,q_s.
\end{eqnarray}
The complete result for $\psi^{(5)}$ is rather involved. To 
simplify the presentation we neglect terms that have vanishing 
matrix elements in transitions to flavour-non-singlet mesons. 
In this case we only need 
\begin{eqnarray}
\label{psi5}
\psi^{(5)} &=&
\frac{1}{in_+\partial}\,g\Slash A_{\perp hc}^{(3)}\,
  \frac{\slash n_+}{2}\,\xi_c
-\frac{1}{in_-\partial}\,\bigg((i\Slash D_{\perp c}+
  g\Slash A_{\perp s}+m) 
  \,\frac{1}{in_+\partial}\,g\Slash A_{\perp hc}^{(3)}\nonumber\\[0.2cm]
&&+\,g\Slash A_{\perp hc}^{(3)}\,
  \frac{1}{in_+\partial}\,(i\Slash D_{\perp c}+
  g\Slash A_{\perp s}-m)\bigg)\,\xi_c
-\frac{1}{in_-\partial}\,g \nm A_{hc}^{(5)}\,\xi_c+\ldots, 
\end{eqnarray}
and the ellipses stand for all the other terms. 
These results have simple interpretations in terms of trees 
of hard-collinear fields. For instance, the second structure in 
$\psi^{(3)}$ comes from the leading term in the solution for $\xi_{hc}$ 
and describes $\xi_{hc}\to A_{\perp c}q_s$ at the end 
of a branch of a hard-collinear tree. Terms involving the hard-collinear 
gluon describe more complicated trees initiated by a quark field, where 
a hard-collinear gluon is emitted, which then splits into 
collinear and soft fields according to the solution for the gluon field 
given below. In a general gauge, arbitrarily complicated trees with 
external $\np A_c$ and $\nm A_s$ fields exist at a given power 
of $\lambda$. The advantage of light-cone gauge is that every branching 
costs factors of $\lambda$. The structure of general trees can be recovered 
from gauge invariance. 

We note that $\psi^{(2)}$ has collinear quark number +1, while 
$\psi^{(3)}$ has collinear quark number 0. The solutions for the 
hard-collinear field below show that $A_{\perp hc}^{(3)}$ and 
$\nm A_{hc}^{(5)}$ have collinear quark number $\pm 1$, while 
 $A_{\perp hc}^{(4)}$ and $\nm A_{hc}^{(4)}$ have 
collinear quark number $0$. 
It follows that  $\psi^{(n)}$ has odd (even) collinear quark number 
for even (odd) $n$. This will be useful later, when we discuss 
the matrix elements of the effective 
heavy-to-light transition current. 

\subsubsection{Hard-collinear gluon field}
\label{subsec:intgluon}

To complete the tree-level matching we consider the Lagrangian
\begin{equation}
\label{startL}
-\frac{1}{4} \,F^2+{\cal L}_{\rm quark}
\end{equation}
with ${\cal L}_{\rm quark}$ the $\SCETI$ quark Lagrangian, in which 
the hard-collinear quark field is eliminated by its tree-level 
solution. We substitute 
$A=A_c+A_s+A_{hc}$ in the Yang-Mills Lagrangian and drop terms not 
allowed by momentum conservation as we did before for the quark
Lagrangian. Integrating out the hard-collinear gluon field at tree level 
requires to solve the classical field equation. This is non-linear, 
but the hard-collinear gluon self-interactions turn out to be 
power-suppressed, so the field equation can be solved as an expansion 
in $\lambda$.
With the ansatz $A_{\perp hc}=\sum_{n=2}^\infty 
A_{\perp hc}^{(n)}$, 
$\nm A_{hc}=\sum_{n=2}^\infty \nm A_{hc}^{(n)}$ 
 ($n$ denotes the power of $\lambda$), we find that the 
leading terms have $n=3$ for $A_{\perp hc}$, 
and $n=4$ for $\nm A_{hc}$. 

The calculation gives 
\begin{eqnarray}
A_{\perp hc}^{(3)} &=& g \, T^A \,
  \frac{1}{(in_+\partial)(in_-\partial)}\,
 \Big\{ \bar q_s \gamma_\perp  T^A \xi_c + {\rm h.c.} \Big\},
\nonumber\\[0.3em]
n_- A_{hc}^{(4)} &=& \frac{2 g}{in_+\partial} \, \Big[A_{\mu_\perp c}, 
   A^{\mu_\perp}_s \Big],
\nonumber\\[0.2cm]
A_{hc}^{\mu_\perp (4)} &=& \frac{g}{(i\np\partial)( i\nm\partial)}\,\Bigg\{
  i{\cal D}^{\mu_\perp} \!\Big[A_{\nu_\perp c}, A^{\nu_\perp}_s \Big] + 
  i{\cal D}_{\nu_\perp} \!\bigg(\Big[A^{\mu_\perp}_s, A^{\nu_\perp}_c 
    \Big] + \Big[A^{\mu_\perp}_c, A^{\nu_\perp}_s \Big] \bigg)
\nonumber\\[0.2cm]
&& +\,\Big[i\np\partial A_c^{\mu_\perp},
  \frac{2g}{i\np\partial}\Big[A^{\nu_\perp}_c, A_{\nu_\perp s} \Big] \Big]+ 
  \frac{i\np\partial}{2}\Big[A^{\mu_\perp}_s, \nm A_c \Big] + 
  \frac{i\nm\partial}{2}\Big[A^{\mu_\perp}_c, \np A_s \Big] 
\nonumber\\[0.2cm] 
&& +\, \Big[i F^{\nu_\perp \mu_\perp}_c, A_{\nu_\perp s} \Big] + 
  \Big[i F^{\nu_\perp \mu_\perp}_s, A_{\nu_\perp c} \Big] 
\nonumber\\[0.2cm]
&& +\, g T^A\,\bar\xi_c\bigg(\Slash A_{\perp s}\frac{1}{i \np \partial}
  \gamma^{\mu_\perp}T^A+\gamma^{\mu_\perp}T^A\frac{1}{i \np \partial}
  \Slash A_{\perp s}\bigg)\frac{\slash n_+}{2}\,\xi_c
\nonumber\\
&& +\, g T^A\,\bar q_s\bigg(\Slash A_{\perp c}\frac{1}{i \nm \partial}
  \gamma^{\mu_\perp}T^A+\gamma^{\mu_\perp}T^A\frac{1}{i \nm \partial}
  \Slash A_{\perp c}\bigg)\frac{\slash n_-}{2}\,q_s
\Bigg\},
\nonumber\\[0.3em]
n_- A_{hc}^{(5)} &=& -\frac{2}{(i\np\partial)^2}\,\Bigg\{
  i{\cal D}^{\mu_\perp} [i\np\partial A_{\mu_\perp hc}^{(3)}] - 
  g\,\Big[i\np\partial A^{\mu_\perp}_c,  A_{\mu_\perp hc}^{(3)}\Big]
\nonumber \\[0.2cm]
&&+\,2g T^A\bigg\{\bar\xi_cT^A\bigg(\frac{\slash n_+}{2}-
  \frac{1}{i\nm \partial}\,g \Slash A_{\perp c}\bigg) q_s+\mbox{h.c.}
  \bigg\}\!\Bigg\}.
\label{sol-Ahc}
\end{eqnarray}
Here $i\,{\cal D}^\mu {\cal O} = i\,\partial^\mu {\cal O} +g\,
[A_c^\mu+A_s^\mu, {\cal O}]$ denotes the covariant derivative on a 
matrix-valued field. We see that at the lowest order the 
transverse hard-collinear gluon branches into a collinear 
and a soft quark, while $\nm A_{hc}$ splits 
into a collinear and a soft gluon. 
Substituting these expressions into 
(\ref{Qfield}) and (\ref{psifield},\ref{psi5}) 
determines $\psi$ and ${\cal Q}$ 
in terms of soft and collinear fields, and the effective current 
follows.  

\subsubsection{Soft-collinear interactions}

As a by-product of the analysis we find the soft-collinear 
interactions in the $\SCETII$ Lagrangian including interactions 
suppressed with $\lambda^3$. We write
\begin{equation}
{\cal L} =  {\cal L}_s+{\cal L}_c+
{\cal L}_{\rm s-c}^{(2)}+{\cal L}_{\rm s-c}^{(3)}+\ldots.
\end{equation}
Soft and collinear fields are decoupled at leading power with 
${\cal L}_s+{\cal L}_c$ given by (\ref{leadinglag}). 
After using the gauge transformation (\ref{eq:cal-soft}) to
convert all expressions into a manifestly gauge-invariant form, 
the tree-level interaction Lagrangian is 
\beq
{\cal L}_{\rm s-c}^{(2)}
&=&
- \,\bar \xi_c W_c \, \Slash {{\cal A}}_{\perp s} \,
  \frac{1}{in_+\partial} \frac{\slash n_+}{2} \, 
 \Slash {{\cal A}}_{\perp s} \, W_c^\dagger \xi_c
 -\bar q_s Y_s  \, \Slash {{\cal A}}_{\perp c} \,
  \frac{1}{in_-\partial} \frac{\slash n_-}{2} \,  
 \Slash {{\cal A}}_{\perp c}  \, Y_s^\dagger q_s
\nonumber \\[0.3em]
&&+ \,\frac{g^2}{2}  \Big\{\bar q_s Y_s \, \gamma_{\perp\mu} \, T^A 
     \, W_c^\dagger \xi_c + {\rm h.c.} \Big\}
  \frac{1}{(in_+\partial)(in_-\partial)}
\Big\{\bar q_s Y_s \, \gamma^{\mu}_\perp \, T^A 
     \, W_c^\dagger \xi_c + {\rm h.c.} \Big\}  
\nonumber \\[0.3em]
&& + \,\frac{1}{g^2} \, {\rm tr} \Big( 
    [{\cal A}_{\mu_\perp c},{\cal A}_{\nu_\perp s}]
    [{\cal A}^{\mu_\perp}_{c},{\cal A}^{\nu_\perp}_{s}]
   \Big)
+ \frac{1}{g^2} \, {\rm tr} \Big(
    [{\cal A}_{\mu_\perp c},{\cal A}_{\nu_\perp c}]
    [{\cal A}^{\mu_\perp}_{s},{\cal A}^{\nu_\perp}_{s}]
   \Big)
\nonumber \\[0.3em] 
&& + \,\frac{1}{g^2} \, {\rm tr} \Big(
    [{\cal A}_{\mu_\perp c},{\cal A}_{\nu_\perp s}]
    [{\cal A}^{\mu_\perp}_{s},{\cal A}^{\nu_\perp}_{c}]
   \Big)
- \frac{1}{g^2} \, {\rm tr} \Big( 
    [{\cal A}_{\mu_\perp c},{\cal A}^{\mu_\perp}_{s}]
    [{\cal A}_{\nu_\perp c},{\cal A}^{\nu_\perp}_{s}]
   \Big).
\label{eq:L8final}
\eeq
In soft-collinear interactions $\int d^4x \,{\cal L}(\phi_c,\phi_s)$ 
we must count $d^4 x$ as $1/\lambda^6$, since $\nm x \sim 1$ is determined 
by the variation of collinear fields, $\np x\sim 1/\lambda^2$ by the 
variations of soft fields, and $x_\perp\sim 1/\lambda^2$, so the vertices 
above give $\lambda^2$ suppression. ${\cal L}_{\rm s-c}^{(3)}$ is lengthy 
and will not be presented here. Recall that in products of soft and 
collinear fields all fields are 
multipole-expanded according to (\ref{eq:multipole}), i.e.\ 
soft fields are taken at $x_-$ and collinear fields at $x_+$ in 
the above interaction terms. This 
implies that $1/(i\np\partial)$ commutes with soft fields, and 
$1/(i\nm\partial)$ commutes with collinear fields. Note that 
in (\ref{eq:L8final}) the $i\epsilon$-prescription of 
the $1/(i n_\mp\partial)$ and the integration path in the 
Wilson lines have to be adjusted according to whether 
one describes incoming or outgoing soft or collinear particles.

The Lagrangian is symmetric under interchange of soft and 
collinear fields together with $n_+ \leftrightarrow n_-$, 
because soft and collinear fields can be transformed into each 
other by a large  Lorentz-boost \cite{Hill:2002vw}.
The existence of the interactions in the first line has already
been mentioned in \cite{Hill:2002vw}. 
The gluon self-interactions in the third and fourth line  could be
written more compactly using the Jacobi identity, but we prefer the 
above symmetric version.

We should emphasize that the soft-collinear interactions do not contribute 
to the matrix elements that define heavy-to-light form 
factors as explained above.

\subsubsection{Effective current}
\label{tree_current}

We now return to the task of determining the $\SCETII$ heavy-to-light 
current at tree level. The current follows from inserting 
the expression for $\psi$ from 
(\ref{psifield},\ref{psi5}) and for the hard-collinear gluon field from
(\ref{sol-Ahc}) into (\ref{jj},\ref{Qfield}). We write the result of this as 
\begin{equation}
J(x)=e^{-im_b v x}\,\left[J_{\rm eff}^{(0)}(x)+J_{\rm eff}^{(1)}(x)+
J_{\rm eff}^{(2)}(x)+J_{\rm eff}^{(3)}(x)+\ldots\right]
\end{equation}
with the superscript indicating the suppression by powers of $\lambda$
relative to the leading term $J_{\rm eff}^{(0)}$. Restoring the gauge 
symmetry the first two terms in the expansion read
\begin{eqnarray}
J_{\rm eff}^{(0)} &=&  \bar \xi_c W_c\, \Gamma \, Y_s^\dagger h_v,
\nonumber\\
J_{\rm eff}^{(1)} &=&  \bar q_s Y_s \, \Slash {{\cal A}}_{\perp c}\, 
 \frac{1}{in_- \overleftarrow \partial} \, \frac{\slash n_-}{2} \, 
 \Gamma \, Y_s^\dagger h_v.
\label{J0J1}
\end{eqnarray}
For $\langle \pi|J_{\rm eff}^{(n)}|\bar B\rangle$ to be non-zero, 
the collinear fields of the operator product must have the quantum numbers 
of the pion, and the soft fields those of the $\bar B$ meson. Hence 
$J_{\rm eff}^{(0)}$ gives no contribution, as already discussed 
earlier. $J_{\rm eff}^{(1)}$ does have the correct collinear and soft 
quantum numbers, except the collinear fields 
(namely $\Slash {{\cal A}}_{\perp c}$) are colour-octet, so the 
matrix element is again zero. The next term in the expansion 
is (in light-cone gauge)
\begin{equation}
J_{\rm eff}^{(2)} = \bar\psi^{(4)}\Gamma h_v -
\frac{1}{\nm v}\,\bar\xi_c\, \Gamma  
\frac{\slash n_-}{2 m_b}\,
g\Slash A_{\perp c} h_v, 
\end{equation}
which carries odd collinear quark number, implying zero matrix element. 

This leaves 
\begin{equation}
\label{j3}
J_{\rm eff}^{(3)} = \bar\psi^{(5)}\Gamma h_v -
\frac{1}{\nm v}\,\bar\psi^{(3)} \Gamma  
\frac{\slash n_-}{2 m_b}\,
g\Slash A_{\perp c} h_v -
\frac{1}{\nm v}\,\bar\xi_c \, \Gamma  
\frac{\slash n_-}{2 m_b}\,
g\Slash A_{\perp hc}^{(3)} h_v,  
\end{equation}
as the leading effective current. Substituting $A_{\perp hc}^{(3)}$ from 
(\ref{sol-Ahc}) we obtain the third term in the gauge-invariant 
representation
\begin{equation}
\label{thirdterm}
 -\left(\frac{g^2}{(in_+\partial)(in_-\partial)}\,
\bar q_s Y_s\, \gamma^{\mu_\perp} T^A W_c^\dagger\xi_c\right)
\frac{1}{\nm v}\,\bar\xi_c W_c \, \Gamma  
\frac{\slash n_-}{2 m_b}\,\gamma_{\mu_\perp}T^A Y_s^\dagger h_v.
\end{equation}
This can be identified with the tree diagram, where a transverse gluon is 
exchanged between the heavy quark and the spectator quark. The 
inverse differential operator is the hard-collinear gluon propagator 
in position space. With $\psi^{(3)}$ from (\ref{psifield}) the second 
term on the right-hand side of (\ref{j3}) has collinear field content 
$A_c$ or $A_c A_c$, which does not contribute to the matrix element, 
when the light meson is a flavour-non-singlet. The first term 
$\bar\psi^{(5)}\Gamma h_v$, which corresponds to trees rooted in the 
hard-collinear quark field at the current vertex, is by far the most 
complicated as seen from the expression for $\psi^{(5)}$ in (\ref{psi5}). 
Although straightforward, it is not 
instructive to give the explicit, gauge-invariant form of this term. 
However, an important observation is that the field content of this 
term is not just $\bar\xi_c\xi_c\,\bar q_s h_v$, but also 
\begin{equation}
\label{j3fields}
\bar\xi_c A_{\perp c}\xi_c\,\bar q_s h_v \quad \mbox{and} 
\quad \bar\xi_c \xi_c\,\bar q_s A_{\perp s} h_v. 
\end{equation}
Eq.~(\ref{psi5}) shows that these terms corresponding to leading-power
tree diagrams with external gluons all originate from 
$\bar\xi_{hc}^{(5)} \hspace*{0.04cm}\Gamma h_v$. (To see this, project 
(\ref{psi5}) with $(\slash n_-\slash n_+)/4$.) On the other hand, 
the remaining projection 
\begin{eqnarray}
\bar \eta^{(5)}_{hc}\hspace*{0.04cm}\Gamma  h_v
&=& \bar \xi_c \frac{\slash n_+}{2}
\,g\Slash A_{\perp hc}^{(3)} 
\left( i n_+ \overleftarrow{\partial}\right)^{-1}
\Gamma h_v
\nonumber\\
&=& \left( \frac{g^2}{(in_+\partial)(in_-\partial)} \,
\bar q_s Y_s \gamma^{\mu_\perp} T^A W_c^\dagger \xi_c\right)
\bar \xi_c W_c \frac{\slash n_+}{2} 
\left( i n_+ \overleftarrow{\partial}\right)^{-1}
\gamma_{\mu_\perp}T^A \Gamma\, Y_s^\dagger h_v
\end{eqnarray}
involves only four-quark operators similar to (\ref{thirdterm}). 
This observation will be crucial for the all-order 
factorization proof in Section~\ref{sec:formfactor}, where we 
define the universal form factor as the matrix element of 
$\bar\xi_{hc}^{(5)} \hspace*{0.04cm}\Gamma h_v$ and show that 
the remaining terms involving four-quark operators factorize into 
a conventional hard-scattering term. 

Summarizing, we conclude that the leading effective current 
is $J_{\rm eff}^{(3)}\sim\lambda^8$, so 
\begin{equation}
\langle \pi(p')|J(0)|B(p)\rangle \approx 
\langle \pi(p')|J_{\rm eff}^{(3)}(0)|B(p)\rangle  
\sim \lambda^3 \sim 
m_b \left(\frac{\Lambda}{m_b}\right)^{3/2}, 
\end{equation}
where we restored the correct dimensions by inserting factors of 
$m_b$.\footnote{Recall that for power counting we often set $m_b=1$, 
so that $\lambda$ counts factors of $\Lambda^{1/2}$.} This reproduces 
the well-known heavy quark scaling of the heavy-to-light 
form factor at large recoil. However, here the scaling has been
derived without recourse to any assumptions on the endpoint behaviour 
of the light-cone distribution amplitude. This result has also been 
obtained in \cite{Bauer:2002aj} in the context of $\SCETI$. 
However, since the matching of external lines with hard-collinear momentum 
to collinear lines that overlap with the meson state has not been 
considered there, the conclusion relied on the implicit 
assumption that the power counting of $\SCETI$ can be applied. 
We may also note that the form factors have an expansion in 
$\lambda^2\sim \Lambda_{\rm QCD}/m_b$ 
once the overall scaling is taken out. This has 
been suggested in \cite{Hill:2002vw} on grounds of the momentum
scaling of soft and collinear fields. This does not 
completely cover the real situation, since the $\SCETII$ interactions 
and currents do have an expansion in $\lambda$ (and not $\lambda^2$), 
\beq
&&  {\cal L}_{\rm s-c} \ = \  {\cal L}_{\rm s-c}^{(2)} + 
{\cal L}_{\rm s-c}^{(3)} + \ldots \ ,
\qquad J_{\rm eff} \ = \ J_{\rm eff}^{(0)} + J_{\rm eff}^{(1)} + \ldots.
\eeq
However, the only source for odd powers of $\lambda$ are the soft 
quark fields, $q_s$ and $h_v$. Since collinear (and soft) quark 
number is conserved by the leading-power action, 
the combinations of ${\cal L}_{\rm s-c}$ and $J_{\rm eff}$ that can 
have non-zero matrix elements must always involve an even number 
of collinear quark or anti-quark fields. Therefore power-corrections
to exclusive observables in $\SCETII$ are indeed given in terms of 
$\Lambda/m_b$.

\subsubsection{Matrix element of the effective current}

The matrix elements of the non-local operators in 
$J_{\rm eff}^{(3)}$ take a more familiar 
form in terms of convolutions with light-cone distribution amplitudes, 
when the factors $1/(i n_\pm \partial)$ are replaced 
using the identities 
\begin{equation}
\label{shift}
\frac{1}{i n_+\partial+i\epsilon}\phi(x) = -i \int^0_{-\infty} 
ds\,\phi(x+s n_+),\quad  
\frac{1}{i n_-\partial-i\epsilon}\phi(x) = i \int_0^{\infty} 
dt\,\phi(x+t n_-).  
\end{equation}
Since all fields are multipole-expanded according to 
(\ref{eq:multipole}), the operator $1/(i \np \partial)$ commutes with 
soft fields since $[x+s \np]_-=x_-$ (see (\ref{xpm}) for the 
definition of $x_\pm$), and  $1/(i \nm \partial)$ commutes with 
collinear fields since $[x+t \nm]_+=x_+$. Any operator can therefore 
be written as a convolution in variables $s_i$, $t_j$ of a product of 
collinear fields with arguments 
$x+s_i \np$ and a product of soft fields with arguments 
$x+t_j \nm$. We exemplify this for (\ref{thirdterm}) and 
then discuss some important general features of the 
matrix element of $J_{\rm eff}^{(3)}$.

Applying (\ref{shift}) the matrix element of (\ref{thirdterm}) turns into 
\begin{eqnarray}
M &\equiv& -g^2\int_{-\infty}^0 ds \int^{\infty}_0 dt \,\langle \pi(p')|
(\bar q_s Y_s)(t \nm)\, \gamma^{\mu_\perp} T^A (W_c^\dagger\xi_c)(s\np)
\nonumber\\
&&\hspace*{1cm}\times \frac{1}{\nm v}\,(\bar\xi_c W_c)(0) \, \Gamma  
\frac{\slash n_-}{2 m_b}\,\gamma_{\mu_\perp}T^A (Y_s^\dagger h_v)(0)
|\bar B(p)\rangle 
\nonumber\\
&=& \frac{g^2 C_F}{N_c}  \,[\gamma^{\mu_\perp}]_{\beta\beta'} \,
\frac{1}{\nm v}\,\Big[\Gamma\frac{\slash n_-}{2 m_b}\,
\gamma_{\mu_\perp}\Big]_{\alpha\alpha'}  
\int_{-\infty}^0 ds  \,\langle \pi(p')|(\bar\xi_{c\alpha} W_c)(0)
\, (W_c^\dagger\xi_{c\beta'})(s\np)|0\rangle 
\nonumber\\[0.2cm]
&& \hspace*{1cm}\times 
\int^{\infty}_0 dt \,\langle 0|(\bar q_{s\beta} Y_s)(t \nm)\, 
(Y_s^\dagger h_{v\alpha'})(0)|\bar B(p)\rangle,
\label{mex1}
\end{eqnarray}
where in the transition to the second line we performed the colour-singlet 
projection on the collinear and soft field products. 
The two matrix elements
define the light-cone distribution amplitudes of the pion and the $B$ 
meson, 
\begin{eqnarray}
&&\langle\pi(p')|
(\bar \xi_{c\alpha} W_c)(s\np)(W_c^\dagger \xi_{c \beta^\prime})(0)
|0\rangle = 
\frac{if_\pi}{4}\,\np p^\prime \left(\frac{\slash n_-}{2}\gamma_5
\right)_{\!\beta^\prime\alpha}\,\int_0^1 du\,
e^{iu s \np p^\prime}\,\phi_\pi(u),
\nonumber\\[0.2cm]
&& \langle 0|(\bar q_{s\beta} Y)(t\nm) (Y^\dagger h_{v \alpha^\prime})(0) 
|\bar B(p)\rangle 
\nonumber\\
&& \hspace*{2.5cm} = -\frac{if_B m_B}{4} \left(
\frac{1+\slash v}{2}\left[\nm v\,\slash n_+\,\tilde \phi_{B+}(t)+
\np v\,\slash n_-\,\tilde \phi_{B-}(t)\right]
\gamma_5\right)_{\!\alpha^\prime\beta},
\label{distamps}
\end{eqnarray}
where $f_\pi=131\,$MeV is the pion decay constant, $f_B$ is the $B$
decay constant and 
\begin{equation}
\tilde\phi_{B\pm}(t)\equiv\int_0^\infty d\omega\,e^{-i\omega t}\,
\phi_{B\pm}(\omega).
\end{equation}
This allows us to rewrite (\ref{mex1}) as 
\begin{eqnarray}
M &=&  \frac{g^2 C_F}{N_c} \,\frac{f_B m_B}{4}
\frac{f_\pi}{4}\,\np p'\,\mbox{tr} \left(\frac{1+\slash v}{2}
\,\slash n_+\gamma_5\,\gamma^{\mu_\perp}\frac{\slash n_-}{2}\gamma_5
\Gamma\frac{\slash n_-}{2 m_b}\,\gamma_{\mu_\perp}\right)
\nonumber\\
&& \hspace*{1cm}\times  \int_0^1 du\,\phi_\pi(u) 
\int_{-\infty}^0 ds \,e^{i (1-u) s \np p^\prime} 
\int_0^\infty d\omega\,\phi_{B+}(\omega) 
\int^{\infty}_0 dt \,e^{-i \omega t}
\nonumber\\[0.2cm]
&=&  \frac{g^2 C_F}{N_c} \,\frac{f_B m_B}{4}
\frac{f_\pi}{4}\,\mbox{tr} \left(\frac{1-\slash v}{2}
\,\slash n_+\slash n_-
\Gamma\frac{\slash n_-}{2 m_b}\right)
\int_0^1 du\,\frac{\phi_\pi(u)}{1-u} 
\int_0^\infty \frac{d\omega}{\omega}\,\phi_{B+}(\omega), 
\end{eqnarray}
which is the desired representation in terms of 
convolutions with light-cone distribution 
amplitudes. 

\begin{figure}[t]
\vspace{1em}
   \centerline{\epsfig{file=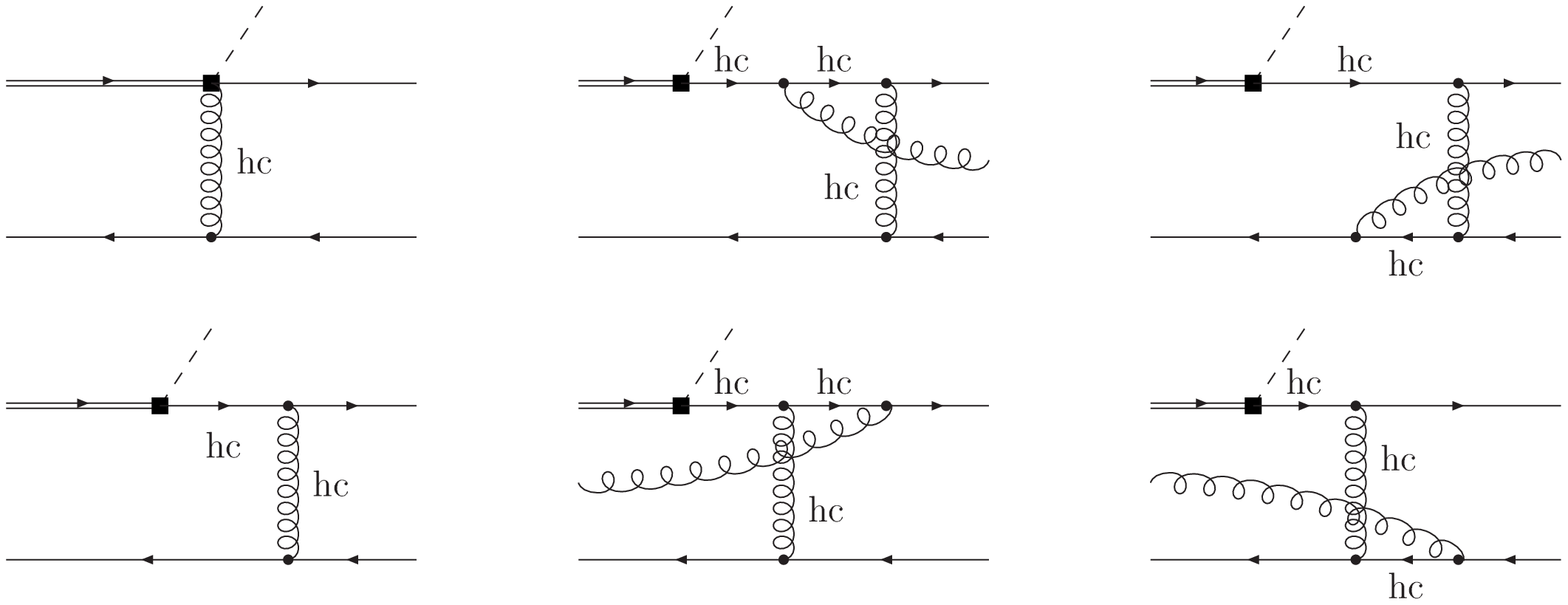, angle=90,
       width=0.9\textwidth, bb = 260 120 500 735}}
\vspace{1em}
\centerline{\parbox{14cm}{\caption{\label{fig:trees} 
Tree diagrams that contribute to the form factor at 
leading power in light-cone gauge. Incoming lines are soft and 
outgoing lines are collinear.}}}
\end{figure}

With the standard assumption that $\phi_\pi(u)$ vanishes 
near the endpoints $u=0,1$ and that $\phi_{B+}(\omega)\to 0$ as 
$\omega\to 0$, the integrals converge in this particular case. But this 
is not true in general. The terms in $\bar\psi^{(5)}\Gamma h_v$ 
contain more powers of $1/(i n_\pm\partial)$. An additional factor 
of $1/(i\nm \partial)$ converts to a factor $1/\omega$ in the convolution, 
and a factor $1/(i\np \partial)$ results in $1/u$, $1/(1-u)$ or 1, 
depending on which fields it operates on. The convolution integrals 
may then be divergent. Furthermore we note that the terms 
(\ref{j3fields}) with the additional collinear or soft gluon fields 
result in contributions from three-particle light-cone distribution 
amplitudes in the pion or the $B$ meson, respectively. The existence 
of endpoint divergences, which prevents the application of the standard 
hard-scattering formalism to the heavy-to-light form factors 
has been known before, but the contribution of three-particle 
amplitudes at leading power is a new and perhaps unexpected 
result. Interestingly, this also appears to happen in QCD sum rule 
calculations of the form factor \cite{Ball:2003bf}. In 
Figure~\ref{fig:trees} we show a sample of tree diagrams contained 
in $\bar\psi^{(5)}\Gamma h_v$, which are not power-suppressed. 
  
\subsubsection{Extension to photons}

We briefly sketch the extension of our results to the 
$B\to\gamma$ transition form factors. These have been studied 
intensively recently  
\cite{Korchemsky:1999qb,Descotes-Genon:2002mw,Lunghi:2002ju,Bosch:2003fc},
since in $B\to\gamma l\nu$ decay theoretical issues related to 
the light-cone structure of the $B$ meson and factorization in 
exclusive $B$ decays 
can be studied without the interference of hadronic final 
state effects, at least in first approximation. 

The electromagnetic coupling is taken into account by substituting
\begin{equation}
  g A_c \to   gA_c^A T^A + e {\cal Q} A_{(\gamma)},
\label{eq:photon-replace}
\end{equation}
in the expressions for the effective Lagrangian and heavy-to-light 
transition current. Here $A_{(\gamma)}$ denotes a collinear photon
field, and ${\cal Q}$ the electric charge operator. 
In the following we always consider the physical 
transverse polarizations of the photon so that $n_\pm A_{(\gamma)}=0$,
and expand to first order in the electromagnetic coupling constant. 

The leading order effective Lagrangian now contains the standard 
electromagnetic coupling $e \bar\xi_c {\cal Q} \Slash A_{\perp
  (\gamma)} \xi_c$ to collinear quarks. This allows unsuppressed
interactions of photons with any number of collinear quark and gluons 
fields. As a consequence matrix elements $\langle \gamma(p')|{\cal O}|
\bar B(p)\rangle$  are non-zero even if ${\cal O}$ contains no 
photon field, making the photon behave similar to a vector meson. 
Such terms are usually referred to as related to the ``hadronic 
structure of the photon''. Hence all the terms at 
order $\lambda^3$ in the
analysis of transitions to mesons can be taken over to the $\bar
B\to\gamma$ transition. 

However, there exist additional terms involving the collinear 
photon field. Such terms are referred to as ``direct photon''
contributions. The corresponding operators ${\cal O}$ contain 
the photon field and soft fields, but no collinear quark and gluon
fields. The most important modification of the analysis for mesons
arises due to the presence of an additional term
\begin{equation}
J^{(1)}_{{\rm eff},(\gamma)} = e e_q \,
   \bar q_s Y_s \, \Slash A_{\perp (\gamma)}\, 
 \frac{1}{in_- \overleftarrow \partial} \, \frac{\slash n_-}{2} \, 
 \Gamma \, Y_s^\dagger h_v 
 = e e_q \,
   \bar q_s \, \Slash A_{\perp (\gamma)}\, 
 \frac{1}{in_- \overleftarrow D_s} \, \frac{\slash n_-}{2} \, 
 \Gamma \, h_v \end{equation}
in the first order current, which has a non-vanishing 
$\langle \gamma|\ldots |\bar B\rangle$ matrix element. 
We therefore conclude that 
\begin{equation}
\langle \gamma(p')|J(0)|\bar B(p)\rangle  \sim \lambda \sim 
m_b \left(\frac{\Lambda}{m_b}\right)^{1/2},   
\end{equation}
i.e. the direct photon contribution is a factor of $m_b/\Lambda_{\rm QCD}$
larger than the hadronic contribution. 
The operator $J^{(1)}_{{\rm eff},(\gamma)}$ is the tree-level 
equivalent of the operator written down in \cite{Bosch:2003fc}. 
Allowing for a non-trivial coefficient function of the operator 
due to radiative corrections, and writing the operator in convolution form 
as discussed above, we reproduce the factorization 
theorem for $B\to\gamma l \nu$ decay at leading order in the $1/m_b$ 
expansion in the form given in \cite{Bosch:2003fc}. 

There exist further direct photon contributions, but they are all 
suppressed by at least two powers of $\lambda$, hence 
they are of the same order as 
the hadronic terms. For instance, photon radiation off the heavy quark
field is obtained from substituting 
(\ref{eq:photon-replace}) into the effective heavy quark field 
${\cal Q}$ in (\ref{Qfield}), which generates the additional 
term 
\begin{equation}
- e e_b \,  \frac{1}{n_- v}\,\bar q_s\,\Gamma \, \frac{\slash n_-}{2m_b}
   \,   \Slash A_{\perp(\gamma)} \, h_v
\end{equation} 
in $J^{(3)}_{\rm eff}$. Hence the direct photon coupling to the 
heavy quark also constitutes a $1/m_b$ correction to the leading 
transition form factor. 

Giving the complete set of $1/m_b$ corrections does not provide 
further insight into the structure of the form factor. In general, 
the issues related to factorization for the $B\to \pi$ form factor 
at leading power, such as endpoint divergences and unsuppressed 
contributions from multi-particle distribution amplitudes, are all
relevant to the $1/m_b$ correction for the $B\to\gamma$ form factor. 
However, this also implies that within the soft-collinear effective theory 
framework it is in principle possible to establish a factorization 
formula for the $B\to\gamma$ form factor, which includes power 
corrections. 

\subsection{Hard-collinear loops}
\label{sec:loops}

Tree-level matching revealed that $\SCETII$ operators are non-local 
in two directions with factors in the product of collinear (soft) 
fields displaced in the $\np$ ($\nm$) direction. The effective theory is 
local in the transverse coordinates. In this section we show that this 
structure is preserved beyond tree level.

It is simpler to begin the discussion in momentum space. Matching 
$\SCETII$ to $\SCETI$ beyond tree level means that we consider an 
$N$-loop diagram in $\SCETI$, where all loop momenta $k_i$ are 
hard-collinear. The external momenta consist of a set of collinear 
momenta $p_j$, a set of soft momenta $l_i$, and, in the case of the 
heavy-to-light current, the heavy quark residual momentum $r$. The 
matching calculation is performed by Taylor-expanding the loop 
integrand under the assumption of hard-collinear loop momentum. 
Apart from numerator factors, which are polynomials in the momenta, 
the integrand can only contain massless propagators and factors 
related to the non-locality of $\SCETI$ vertices. The massless 
propagators are expanded as 
\begin{equation}
\frac{1}{(K+L+P)^2} = \frac{1}{K^2+\np K \nm L+\nm K \np P+\np P\nm L} 
+\ldots,
\end{equation}
where $K$ ($L$, $P$) denotes a linear combination of hard-collinear 
(soft, collinear)  momenta. In case of the heavy-to-light current, we 
exploit that the heavy quark line ends at the current vertex, so 
that the residual momentum can be routed such that it never enters a 
massless propagator. The non-locality of $\SCETI$ vertices may supply 
factors of $1/(\np (K+P))$. The only external vectors that enter the 
loop integrand denominator are therefore $(\np p_j) \,\nm/2$ and 
$(\nm l)\,\np/2$, from which we can build the Lorentz-invariants 
$Q_{ij}=(\nm l_i)(\np p_j)$. When the loop integrals are computed in 
dimensional regularization, the dimensionless short-distance 
part can only be of the form 
\begin{equation}
f(Q_{i'j'}/Q_{ij},\ln(Q_{ij}/\mu^2))
\end{equation} 
that is an arbitrary function of ratios of invariants and a polynomial 
of logarithms of the factorization scale. 

The dependence of a matching coefficient on an external momentum 
implies that the corresponding operator is non-local. Here the 
non-locality is due to the fact that the $\np$ component 
of the hard-collinear 
momentum is not larger than the $\np$ component of the external collinear 
momentum, and the $\nm$ component of the hard-collinear momentum is not 
larger than the $\nm$ component of the soft momentum. In position 
space the invariants $Q_{ij}$ correspond to factors 
$1/(i\nm\partial \,i\np \partial)$ acting on certain products of fields. 
It is more convenient to use (\ref{shift}) to represent these factors 
as integrations. The general form of a soft-collinear operator including 
its coefficient function reads
\begin{eqnarray}
&&\int_{-\infty}^\infty ds_1\ldots ds_n
\int_{-\infty}^\infty dt_1\ldots dt_m  
 \,s_i^a t_j^b\,C\Big((s_{i'}t_{j'})/(s_i t_j), \ln(s_i t_j\mu^2)\Big)\,
\nonumber\\[0.2cm]
&&\hspace*{1cm}\times\,
[\phi_{s1}(x+t_1 \nm)\ldots\phi_{sm}(x+t_m\nm)]\,
[\phi_{c1}(x+s_1 \np)\ldots\phi_{cn}(x+s_n\np)].
\end{eqnarray}
We extended the integration limits to infinity, since the precise
limits depend on the details of how the multiple 
inverse derivative operators act on products of fields. The correct
limits are then implemented through theta-functions in $C$.
The two square brackets contain the non-local products of soft and 
collinear fields, respectively, which may contain 
derivatives or Dirac matrices depending on the specific case. 
The coefficient function $C$ is 
dimensionless by construction. It should be noted that the $s_i$ and 
$t_j$ have mass dimension $-1$ and transform as $s_i\to 
\alpha s_i$, $t_j\to \alpha^{-1} t_j$ under the boost transformation 
$\nm\to \alpha\nm$, $\np\to\alpha^{-1}\np$, under which the 
effective theory must remain invariant. Knowing the boost transformation 
of the $\SCETI$ operator implies constraints on the possible 
occurrence of ``free'' factors of $s_i$ and $t_j$, indicated above as 
$s_i^a t_j^b$. The Fourier transforms of the 
coefficient functions defined here appear as hard(-collinear) 
scattering kernels in factorization theorems. In general, 
all interaction and operator vertices in $\SCETII$ 
take the form of convolution kernels.

\section{Factorization of heavy-to-light form factors}
\label{sec:formfactor}

We now turn to the proof of the factorization formula (\ref{ff1}) for 
$B$ meson transition form factors to an energetic light meson. The
light meson is assumed to be a flavour non-singlet pseudoscalar, such 
as the pion, but a similar result holds for vector mesons as discussed 
below.  
 
The three independent form factors for $\bar{B}$ decays into a pion are  
defined by the following Lorentz decompositions of bilinear quark  
current matrix elements: 
\begin{eqnarray} 
&& \langle \pi(p')|\bar u\, \gamma^\mu b |\bar{B}(p)\rangle = 
f_+(q^2)\left[p^\mu+p^{\prime\,\mu}-\frac{m_B^2-m_\pi^2}{q^2}\,q^\mu\right] 
+f_0(q^2)\,\frac{m_B^2-m_\pi^2}{q^2}\,q^\mu, 
\\ 
&& \langle \pi(p')|\bar u \, \sigma^{\mu\nu} q_\nu b|\bar{B}(p) \rangle = 
\frac{i f_T(q^2)}{m_B+m_\pi}\left[q^2(p^\mu+p^{\prime\,\mu})- 
(m_B^2-m_\pi^2)\,q^\mu\right], 
\label{ffdef} 
\end{eqnarray} 
where $m_B$ is the $B$ meson mass, $m_\pi$ the pion mass, and $q=p-p'$.  
The result to be shown is that for $i=+,0,T$, and for $q^2\ll m_B^2$ 
(such that the energy of the pion $E\gg\Lambda$),   
\begin{equation} 
\label{fact2} 
f_i(q^2) = C_i(q^2)\,\xi_\pi(q^2) + \int_0^\infty \!d\omega \int_0^1 \!du\,  
T_i(q^2;\omega,u) \,\phi_{B+}(\omega)\,\phi_\pi(u), 
\end{equation} 
up to corrections of order $\Lambda/m_B$, and that the following holds: 
\begin{itemize} 
\item[1.] The coefficients $C_i$ and $T_i$ are dominated by 
  short-distance physics at the scales $m_B$ and $\sqrt{m_B\Lambda}$  
  and have expansions in $\alpha_s$. 
\item[2.] The form factor $\xi_\pi$ is universal, i.e. independent 
  of $i=+,0,T$. ($\phi_{B+}$ and $\phi_\pi$ are light-cone  
  distribution amplitudes defined later.) 
\item[3.] The integrations over $\omega$ and $u$ converge. In 
  particular, there are no ``endpoint singularities''.  
\end{itemize} 
The proof proceeds in three steps: we first determine the  
$\SCETI$ operators, which give leading-power contributions to the form 
factors after matching them to $\SCETII$. We then define the universal 
form factor $\xi_\pi$. Finally, we show that the remainder  
$f_i-C_i\xi_\pi$ matches only on a particular type of  
$\SCETII$ operator, which factorizes into light-cone distribution  
amplitudes, and for which the convolution integrals are convergent. 
The strategy which we apply here is to avoid having to deal with 
endpoint divergences by applying a definition of $\xi_\pi(q^2)$ that subsumes 
all these effects. This comes at the price of not factorizing 
$\xi_\pi(q^2)$ into its hard-collinear, collinear and soft 
subprocesses. However, this is quite sufficient, since the main 
outcome of the formula (\ref{fact2}) that the three independent 
form factors reduce to one plus a hard-scattering term does 
not require the factorization of $\xi_\pi(q^2)$.
 
\subsection{Matching on $\SCETI$: Integrating out hard modes} 
 
At leading order the flavour-changing QCD currents  
$\bar \psi \hspace{0.04cm}\Gamma_i Q$ match to operators 
\begin{equation} 
\label{op41}
\bar\xi_C W_C \Gamma^\prime_j h_v 
\end{equation} 
in $\SCETI$, where $\Gamma^\prime_j$ is a basis of Dirac matrices. 
To avoid confusion with the label ``c'' for collinear fields 
in $\SCETII$ we use the label ``C'' to denote collinear quantities 
in $\SCETI$, when they describe hard-collinear {\em and} collinear 
modes. The operators (\ref{op41})  
scale with $\lambda^4$. The QCD operators have been matched to SCET  
at tree level up to order $\lambda^6$ \cite{BCDF}, and the complete  
operator basis is known at order $\lambda^5$ \cite{Pirjol:2002}.  
The $\lambda$ scaling here is derived from counting the $C$-fields as  
hard-collinear, since this gives the smallest power, and covers the 
general case. 
 
The tree-level matching calculation  
showed that the leading  $\SCETII$ operators that have non-vanishing  
$\langle\pi|\ldots|\bar B\rangle$ matrix 
elements scale with $\lambda^8$. (The form factor  
then scales as $\lambda^3\sim  1/m_b^{3/2}$ as expected.) However, it 
is not necessary to enumerate the complete basis of current  
operators in $\SCETI$ up to order $\lambda^8$, since most of these 
operators contribute to the $\langle \pi|\ldots|\bar B\rangle$ matrix 
element only through time-ordered products with power-suppressed  
$\SCETI$ interactions. Rather than giving the general form of the 
$\SCETI$ current after integrating out the hard modes, we restrict  
our attention to those operators that contribute to the leading-power 
form factor. To identify these operators, we distinguish 
hard-collinear and collinear fields in the following.   
 
For $\SCETI$ operators that scale as $\lambda^n$, the power $n$ is 
never smaller than the scaling 
of the product of fields involved, because there are  
no $1/(i\nm \partial)$ factors in $\SCETI$. We therefore first list  
the operators by their field content, using the  
gauge $\np A_c=\np A_{ hc} =\nm A_s=0$. The 
gauge-invariant representation of the operators will be constructed 
later.  
 
Consider first operators without hard-collinear fields. To represent 
the quantum numbers of the  
$\bar B$ meson, the fields $\bar q_s h_v$ are needed. Adding 
collinear fields for the outgoing light meson, the only  
operator at order $\lambda^8$ is $\bar{q}_s h_v A_c$, but this cannot 
contribute to the 
$\langle \pi|\ldots|\bar B\rangle$ matrix element, since the collinear 
fields are colour-octet flavour-singlet. The leading purely 
soft-collinear operators with the correct quantum numbers  
arise at order $\lambda^{10}$, the two possibilities reading  
$\bar{q}_s h_v \bar{\xi}_c\xi_c$, $\bar{q}_s h_v A_c A_c$. The  
second operator is flavour-singlet and 
irrelevant for pions.\footnote{To avoid  
heavy notation we do not indicate the flavour of the collinear and 
soft quark fields, which should be clear from the context.} The first  
is generated by the hard contribution to the box graph representing 
gluon exchange from the heavy quark and the light quark to  
the spectator quark. As expected 
this results in a $1/m_b$ power correction to the form factor. 
 
Any relevant $\SCETI$ operator must therefore contain at least one  
hard-collinear field. With the power counting for hard-collinear
fields, $\xi_{hc}\sim \lambda$, $A_{\perp hc}\sim\lambda$, 
$\nm A_{hc}\sim \lambda^2$, we obtain  
\begin{equation} 
\label{oplist} 
\begin{array}{ll} 
\lambda^4 \quad & \bar\xi_{ hc}h_v \\[0.2cm]  
\lambda^5 & \bar\xi_{ hc}A_{\perp{ hc}}h_v\\[0.2cm] 
\lambda^6 & 1)\hspace*{0.2cm}  \bar\xi_c A_{\perp{ hc}}h_v,\quad  
            2)\hspace*{0.2cm} \bar\xi_{ hc}A_{\perp{c}}h_v,\quad 
            3)\hspace*{0.2cm} \bar\xi_{ hc}A_{\perp{s}}h_v,\quad 
            4)\hspace*{0.2cm} \bar\xi_{ hc} \nm A_{\rm 
              hc}h_v,\\[0.2cm]  
          & 5)\hspace*{0.2cm} \bar\xi_{ hc}A_{\perp{ hc}} 
              A_{\perp{ hc}}h_v, 
\end{array} 
\end{equation} 
and so on. We then need to determine the $\lambda$ suppression  
factors incurred when the hard-collinear fields convert into  
soft and collinear fields through time-ordered products.  
We show now that this implies that no operators with hard-collinear  
fields that scale as $\lambda^7$ or $\lambda^8$ in $\SCETI$  
contribute to the leading-power form factors, and that only a few  
of the operators listed above are actually relevant. 
 
To find the suppression factors we inspect the interaction vertices of 
the $\SCETI$ Lagrangian. There exist unsuppressed interactions among the 
hard-collinear modes but any interaction that couples hard-collinear 
to collinear and/or soft modes costs at least a factor of 
$\lambda$ (in the light-cone gauges adopted above).  
It follows that $\SCETI$ operators with hard-collinear 
fields at order $\lambda^8$ contribute to the form factors only at 
sub-leading order. To proceed we note that the $\SCETII$ current  
operators must have the field content  
\begin{equation} 
(\bar{q}_sh_v[\ldots]) \ (\bar\xi_c\xi_c[\ldots])  
\end{equation} 
to match the quantum numbers of the $\bar B$ meson and the pion,  
where the first ellipses denote additional soft fields (gluons or  
quark-antiquark pairs), the second ellipses additional collinear 
fields, and $\bar\xi_c\xi_c$ is flavour non-singlet.  
In particular, the simplest operators are four-quark 
operators.\footnote{In case of flavour-singlet mesons the collinear 
  fields may be gluons only, but we do not consider this case here.}  
Suppose now that $J_7$ is a $\SCETI$ current operator, which scales as 
$\lambda^7$, and contains at least one hard-collinear field. For this 
operator to contribute to the form factor at leading power, there 
should be a time-ordered product  
\begin{equation} 
\label{to7} 
\int d^4 x\,T(J_7,{\cal L}_{\rm int}(x)) 
\end{equation} 
of order $\lambda^8$ that matches onto the required four-quark 
operator. Besides $h_v$ the current $J_7$ can contain only  
$\bar q_s$ or $\bar\xi_c$ (but not both) and no further soft or  
collinear fields, so the question is whether there is a  
$\SCETI$ interaction  
that produces $\bar q_s\xi_c$ or $\bar\xi_c\xi_c$ at the price of only 
a single factor of $\lambda$. According to (\ref{collfinal}) the  
only possible candidate is  
\begin{equation} 
\label{int1} 
{\cal L}_{{\rm int}, \xi q} = \bar q_s g\Slash A_{\perp  hc} \xi_c, 
\end{equation} 
but since this interaction contains a only single hard-collinear 
field, it must occur at the end of a branch of a 
hard-collinear tree. Tree-level matching tells us that the  
hard-collinear gluon that connects to this branch counts as 
$\lambda^3$ rather than $\lambda$, so the time-ordered product  
(\ref{to7}) is at least of order $\lambda^9$. More explicitly,   
\begin{equation} 
\label{top1} 
A_{\perp hc} \int d^4 x \,\bar q_s g\Slash A_{\perp  hc} \xi_c 
\sim\lambda^3, 
\end{equation} 
\unitlength1cm 
\begin{picture}(0,0)(0,0.2) 
 \put(5.9,0.8){\line(0,-1){0.3}} 
 \put(5.9,0.5){\line(1,0){2.6}} 
 \put(8.5,0.5){\line(0,1){0.3}} 
\end{picture} 
\vskip-0.1cm\noindent 
since the contraction integrated over space-time scales as 
$1/\lambda^2$, as the hard-collinear propagator in momentum space.  
Since $A_{\perp hc}$ counted as $\lambda$ when we determined the 
scaling of $J_7$, we see that the conversion of $ A_{\perp hc}$  
into $\bar q_s\xi_c$ costs a factor of $\lambda^2$. It follows  
that no operator that scales with $\lambda^7$ in $\SCETI$ is relevant 
to the form factors at leading power.  
 
We are left with the list (\ref{oplist}), which we now 
reduce. Consider first the leading operator $J_4=\bar\xi_{ hc} 
h_v$. A non-zero matrix element requires at least two suppressed 
interaction vertices, since no single interaction can produce the 
required  $\bar\xi_c\xi_c\bar q_s$ fields. The leading candidate is  
\begin{equation} 
\label{to4} 
\int d^4 x\,d^4y\,T(J_4,{\cal L}_{{\rm int},\xi\xi}(x), 
{\cal L}_{{\rm int},\xi q}(y)) 
\end{equation} 
with ${\cal L}_{{\rm int},\xi q}$ from (\ref{int1}) and  
${\cal L}_{{\rm int},\xi\xi}$ an interaction of the form  
$\bar\xi_c[\ldots]\xi_{ hc}$ from the collinear Lagrangian. The 
second interaction implies a factor of $\lambda^2$ as shown above, but 
it appears that there are order $\lambda$ interactions such as  
$\bar\xi_c A_{\perp, hc} (i\np\partial)^{-1}\partial_\perp \xi_{\rm 
  hc}$ in ${\cal L}_{{\rm int},\xi\xi}$, in which case the time-ordered 
product scales with $\lambda^7$. However, an additional factor of 
$\lambda$ arises, because one can show that in an arbitrary diagram 
contributing to the time-ordered product, there is always an odd 
number of transverse derivatives. Although these scale as $\lambda$ in 
general, at least one of them must act on an external field (in which 
case it scales as $\lambda^2$), since the integrations over 
an odd number of factors of internal transverse momenta vanish. Hence $J_4$  
contributes to the relevant $\lambda^8$ current operators in  
$\SCETII$. Some of the tree-level terms discussed earlier can in fact 
be traced to this time-ordered product.  
 
The time-ordered products needed to generate a non-vanishing  
$\langle \pi|\ldots|\bar B\rangle$ matrix 
element of the operator $J_5=\bar\xi_{ hc}A_{\perp{ hc}}h_v$ are 
the same as for $J_4$. The additional $\lambda$ suppression is not 
present here, because the operator $J_5$ already supplies one 
transverse gluon field, so $J_5$ also belongs to the set of operators  
relevant to the form factors at leading power. However,  
there is no tree-level contribution from $J_5$, since $J_5$ can 
contribute at order $\lambda^8$ only through hard-collinear loops.  
 
Turning to the five operators that scale with $\lambda^6$ in the list  
(\ref{oplist}), we can immediately discard all except the first 
one. For the operators 2) and 3) this follows, because we require the 
same time-ordered products as in (\ref{to4}) to obtain a non-vanishing 
matrix element, but the additional collinear or soft gluon field in 
the operator gives an extra $\lambda^2$ suppression. Similarly, the 
operators 4) and 5) can be dropped by comparing them to  
$J_5$. This leaves $J_6=\bar\xi_{c}A_{\perp{ hc}}h_v$, which  
gives a relevant $\lambda^8$ term through  
\begin{equation} 
\int d^4 x\,T(J_6,{\cal L}_{{\rm int},q\xi}(x)). 
\end{equation} 
This contribution is already present at tree level (third term in 
(\ref{j3})).  
 
Having determined the field structure of the operators, we note  
that there are only three independent Dirac structures between 
$\bar\xi_{c}$ (or $\bar\xi_{ hc}$) and $h_v$, which we choose as  
\begin{equation} 
 \Gamma_j^\prime = \{1,\gamma_5,\gamma^{\mu}_\perp\}. 
\end{equation} 
The only $\SCETI$ currents that we need are therefore of the  
form  
\begin{equation}  
\label{3ops} 
\bar{\xi}_{ hc}\Gamma_j^\prime h_v,\quad  
\bar{\xi}_{ hc}A_{\perp{ hc}}^{\mu} \Gamma_j^\prime h_v, \quad  
\bar{\xi}_c A_{\perp{ hc}}^{\mu}\Gamma_j^\prime h_v. 
\end{equation} 
Any other operator that may appear in the representation of the  
QCD weak currents in $\SCETI$ contributes only power corrections to 
the form factors. In the notation of \cite{BCDF} the three relevant  
operators descend from $J^{(A0)}$, $J^{(A1)}$ and $J^{(B1)}$. 
 
To put the position arguments of the fields and gauge invariance in 
place, we note that fields are multipole-expanded in the position 
space formulation of SCET \cite{BCDF}, and that the (hard-)collinear fields 
are in general separated in the $\np$ direction along the light-cone 
due to the action of $(i\np\partial)^{-1}$. For the following  
gauge-invariant representation it is convenient to 
temporarily return to the formulation of $\SCETI$, where a collinear 
field represents hard-collinear and collinear modes. The multipole 
expansion of soft fields around $x_-=(\np x) \nm/2$ performed in  
\cite{BCDF,Beneke:2002ni} is not appropriate now, because it applies 
to a theory with only hard-collinear and soft modes, where the 
transverse variations of soft fields are always small compared to those 
of the hard-collinear fields. Gauge invariance in $\SCETI$ in 
situations with collinear and hard-collinear modes has not been 
discussed in the literature so far in either of the two formulations 
of SCET (position space, hybrid), and we do not intend to give a 
systematic discussion here. In position space a simple remedy  
appears to be to include the transverse position in the definition of  
$x_-$ and to multipole-expand the soft fields around  
$x_-=(\np x) \nm/2+x_\perp$.   
The gauge transformation of the transverse collinear field in  
Eq.~(10) of \cite{Beneke:2002ni} should then be replaced by 
\begin{equation} 
A_{\perp C}\to U_{C} \, A_{\perp C} \, U^{\dagger }_{C} 
   + \frac{i}{g} \, U_{C} 
   \left[D_{\perp\rm s}(x_-), U^{\dagger }_{C}\right]. 
\end{equation} 
(Remember that the subscript ``C'' includes hard-collinear and  
collinear modes.) Contrary to the construction in \cite{Beneke:2002ni} 
the gauge transformations do not obey homogeneous power counting rules in 
$\lambda$. This is not surprising, since the $\SCETI$ Lagrangian  
has no homogeneous power counting, when  
hard-collinear and collinear modes are represented by a single 
field. The gauge-invariant form of the operators (\ref{3ops})  
is now given by 
\begin{eqnarray} 
O^0_j(s) &\equiv&  (\bar\xi_C W_C)(x+s\np) \Gamma_j^\prime \,h_v(x_-)  
\equiv (\bar\xi_C W_C)_s \Gamma_j^\prime \,h_v  
\label{o0}\\[0.2cm] 
O_j^{1\mu}(s_1,s_2) &\equiv&  
(\bar\xi_C W_C)_{s_1}(W_C^\dagger 
iD_\perp^\mu W_C)_{s_2}\Gamma_j^\prime \,h_v   
\label{o1} 
\end{eqnarray} 
with $D_\perp$ acting on everything to its right.\footnote{The  
interpretation of the subscript ``s'' as denoting a soft field  
or the position argument $x+s\np$ should be clear from the context.}  
In light-cone gauge  
$W_C=1$, and the second operator reduces to $\bar{\xi}_C 
A_{\perp{hc}}^{\mu} \Gamma_j^\prime h_v$ and further $\lambda$  
suppressed terms, which we can drop.  
 
The operators are multiplied by coefficient functions, whose 
tree-level values are corrected by hard loops. Lorentz invariance 
implies that the coefficient functions can depend only on the 
invariants $m_b^2$ and $m_b \nm v \,\np p_k^\prime$ (both of order 
$m_b^2$) with $p_k^\prime$ referring to a set of independent external 
collinear or hard-collinear momenta. The other possible invariants  
are small compared to $m_b^2$ and Taylor-expanded before the 
computation of the loop integral \cite{BCDF}. Invariance under  
boosts, $\nm\to\alpha\nm$, $\np\to \alpha^{-1}\np$, implies the 
combination $\nm v \,\np p_k^\prime$, as can also be verified explicitly 
from the structure of loop integral denominators. The dependence of 
the coefficient function on an external momentum component results in 
non-locality and convolutions in coordinate space.  
If the position argument of the field is $x+s_i\np$,  
we express the coefficient function in terms of 
the dimensionless and boost-invariant convolution variable  
$\hat s_i\equiv s_i m_b/\nm v$.  
 
The QCD currents $\bar\psi \hspace{0.04cm}\Gamma_i Q$ are  
now represented in $\SCETI$  
as  
\begin{eqnarray} 
\label{scet1currents} 
(\bar\psi \hspace{0.04cm}\Gamma_i Q)(x) &=&  
e^{-im_b v \cdot x} \Bigg\{ \sum_j \int  
d \hat s \, \tilde C^{\;0}_{ij}\, (\hat s, \frac{m_b}{\mu})\,O^{0}_j(s) 
\nonumber\\[0.1cm] 
&&\,+ \ \frac{1}{m_b} \ \sum_j \; \int d\hat s_1  
d\hat s_2 \, \tilde C_{ij}^{1\mu}(\hat s_1, \hat s_2, \frac{m_b}{\mu}) \,  
O_j^{1\mu} (s_1,s_2)\Bigg\} + \ldots, 
\end{eqnarray} 
where further terms that do not contribute to the leading-power 
operators after matching to $\SCETII$ are not written explicitly. This 
is the main result of the first step in the factorization proof. The 
factor $1/m_b$ in the second line has been inserted so that the 
coefficient functions are dimensionless. The indices `$ij$' are 
schematic and contain Lorentz indices in general. The coefficient 
functions are boost-invariant Lorentz tensors constructed from  
$n_\mp^\mu$, $g^{\mu\nu}$, $\epsilon^{\mu\nu\rho\sigma}$ and  
$\nm v$.\footnote{Without loss of generality we may adopt a frame 
  where $v_\perp=0$, so $\np v=1/\nm v$.} For instance, if 
$\Gamma_i=i \sigma^{\mu\nu}$ and $\Gamma^\prime_j=\gamma_\perp^\rho$, 
then $\tilde C^{\;0}_{ij}$  contains two independent scalar coefficients 
functions multiplying $(n_{-\mu} g_{\nu\rho}-n_{-\nu} g_{\mu\rho})/\nm 
v$ and  $(n_{+\mu} g_{\nu\rho}-n_{+\nu} g_{\mu\rho}) \nm v$, 
respectively. The reparameterization-invariance constraints on the 
coefficient functions discussed in the literature  
\cite{BCDF,Pirjol:2002,Chay:2002vy} refer to 
these scalar coefficients.  
However, for the following considerations it will not be 
necessary to enumerate all the possible structures in detail.  
The notation (\ref{scet1currents}) also implies that the matrix elements 
of the operators on the right-hand side are taken with respect to the 
full $\SCETI$ Lagrangian, not just the leading-power Lagrangian. For the 
purposes of power counting and renormalization-group evolution it is 
sometimes convenient to make explicit the effect of power-suppressed 
interactions in the form of time-ordered products, however, we shall not 
use this convention here; the time-ordered product operators are part 
of the matrix elements with the full $\SCETI$ Lagrangian.
 
The $\mu$-dependence of the coefficient functions has two origins. One 
is the scale dependence of the QCD weak currents, unless the current 
is conserved. The other is related to the factorization of the hard 
modes and is specific to the effective theory. This dependence 
is compensated by the dependence of the operators $O^{0}_j(s)$,  
$O_j^{1\mu} (s_1,s_2)$ on $\mu$. We implicitly assume a factorization 
scheme (dimensional regularization with minimal subtractions) that 
does not introduce a hard cut-off. Then $O_j^{1\mu} (s_1,s_2)$ does not 
mix into $O^{0}_j(s)$ under renormalization, since it is 
$\lambda$ suppressed relative to  $O^{0}_j(s)$  in $\SCETI$. On the other 
hand the time-ordered products with power-suppressed $\SCETI$ interactions 
included in the matrix element of  $O^{0}_j(s)$ may require  
counterterms proportional to $O_j^{1\mu} (s_1,s_2)$, so the converse is 
not true. Schematically, we have
\begin{equation}
\mu\frac{d}{d\mu}\left(\begin{array}{c} 
\langle O^{0}_j \rangle \\[0.1cm]  
\langle O_j^{1} \rangle
\end{array}\right) = 
\left(\begin{array}{cc} 
* & * \\[0.1cm]
0 & * \end{array}\right)
\left(\begin{array}{c} 
\langle O^{0}_j \rangle \\[0.1cm]  
\langle O_j^{1} \rangle
\end{array}\right)
\end{equation}
where $\langle \ldots\rangle$ denotes taking an appropriate matrix element. 

 In \cite{BCDF} there appeared two operators at tree level at order  
$\lambda$ with fields $\bar\xi_C A_{\perp C} h_v$, which take the expressions  
\begin{equation} 
-\bar\xi_C\,\Gamma_i\,\frac{\slash n_-}{2 m_b}\, 
[i D_{\perp C}^\mu W_C]\,h_v, 
\qquad  
-\bar\xi_C \,i \overleftarrow{D}_{\perp C}^\mu
\frac{1}{i\np \overleftarrow{D}_C} 
\frac{\slash n_+}{2} \Gamma_iW_C\,h_v. 
\end{equation} 
Despite their different appearance both structures can be related to  
(\ref{o1}). If at tree level $\tilde C_{ij}^{1\mu}$ has pieces  
proportional to $\delta(\hat s_1)\delta(\hat s_2)$, then the 
corresponding term in (\ref{scet1currents}) collapses to  
\begin{equation} 
\frac{1}{m_b}\,\bar\xi_C\,[i D_{\perp C}^\mu W_C]\,\Gamma^\prime_j h_v, 
\end{equation} 
of which the first of the two structures is a linear combination.  
Pieces proportional to $\delta(\hat s_1-\hat s_2)$ result in  
\begin{equation} 
\frac{1}{m_b}\int_{-\infty}^0 d\hat s\,(\bar\xi_C \,i
\overleftarrow{D}_{\perp C}^\mu  
W_C)_{s} \Gamma_j^\prime \,h_v   =  
\frac{1}{\nm v}\,\bar\xi_C \,i\overleftarrow{D}_{\perp C}^\mu\frac{1}
{i\np \overleftarrow{D}_C} W_C \,\Gamma_j^\prime \,h_v, 
\end{equation} 
which relates to the second structure.  
The result (\ref{scet1currents}) given above is also  
consistent with the analysis of the general operator 
basis given in \cite{Pirjol:2002}, simplified to the choice  
$v_\perp=0$.  
 
\subsection{Definition of the $C_i\,\xi_\pi$ term} 
 
We define $\xi_\pi(q^2)$ through the matrix elements  
of the $\SCETI$ operators  
$\bar\xi_C\,\Gamma_j^\prime h_v$. Since only $\Gamma_j^\prime=1$ 
does not vanish between $\langle\pi(p')|$ and $|\bar B(p)\rangle$, 
this defines only one function, independent of the original  
Dirac matrix $\Gamma_i$. The precise definition reads\footnote{We do 
  not distinguish $|\bar B(p)\rangle$ from the meson eigenstate  
  $|\hat B_v\rangle$ of the leading order HQET Lagrangian, because the 
  difference is a higher-order effect.} 
\begin{equation} 
\label{xidef} 
\langle\pi(p')|(\bar\xi_C \,W_C \,h_v)(0)|\bar B(p)\rangle \equiv  
2 E\,\xi_\pi(q^2) 
\end{equation} 
with  $E=(\nm v)(\np p')/2=(m_B^2-q^2)/(2 m_B)$  
the energy of the outgoing pion in the $\bar B$ meson rest frame. 
(We neglect $m_\pi^2 \sim \lambda^4$.)  
By definition $\xi_\pi(q^2)$ includes also dynamical effects at  
the scale $\sqrt{m_B\Lambda}$ through hard-collinear effects. In  
(\ref{xidef}) collinear expressions include collinear and 
hard-collinear, and the matrix element is taken with respect to the 
full $\SCETI$ Lagrangian, not just the leading-power Lagrangian 
(in which case the matrix element would vanish). As explained above, 
$\xi_\pi(q^2)$ also depends on a renormalization and factorization 
scale $\mu$, such that $d\xi_\pi(q^2)/d\mu$ may contain a term proportional 
to the matrix element of $O_j^{1\mu} (s_1,s_2)$.
  
The matrix element of the first term on the right-hand side of  
(\ref{scet1currents}) can be expressed in terms of $\xi_\pi(q^2)$,  
since one collinear convolution integral can always be trivially  
done. Choosing $x=0$,  
\begin{eqnarray} 
&&   \langle\pi(p')|\sum_{j} \int  d\hat s \,  
  \tilde C_{ij}^0(\hat s,m_b/\mu) \, O^0_j(s) |\bar B(p)\rangle 
\nonumber\\[0.2cm] 
&&\hspace*{1cm}  =  \int d\hat s \,  
  \tilde C_{i1}^0(\hat s,m_b/\mu)\,\langle\pi(p')|\,e^{i s\np P}\,  
  (\bar\xi_C\,W_C \,h_v)(0)\,e^{-i s\np P} |\bar B(p)\rangle 
\nonumber\\[0.2cm] 
&&\hspace*{1cm}  =  \int d\hat s \,e^{i s\np p^\prime} \, 
  \tilde C_{ij}^0(\hat s,m_b/\mu)\,\langle\pi(p')|\bar\xi_C\,W_C \,h_v 
  |\bar B(p)\rangle 
\nonumber\\[0.2cm] 
&&\hspace*{1cm}  =  2 E\,C_i(E/m_b,m_b/\mu) \,\xi_\pi(q^2)  
\label{matr1}
\end{eqnarray} 
where in the second line $P$ is the momentum operator. In passing to 
the third line, we neglected $p-m_b v$, since this difference  
is important only beyond leading power.  
(The momentum operator in the effective theory  
on $|\bar B(p)\rangle$ gives $p-m_b v$, because the large component  
$m_b v$ is scaled out.) In the last line we used (\ref{xidef}),  
and defined the momentum-space 
coefficient function $C_i(E/m_b,m_b/\mu)$. 
 
A few remarks on the definition of $\xi_\pi(q^2)$ are in order: 
 
(i) The definition is not unique. For instance, it 
can be redefined by a hard-scattering term. This freedom has been used 
in \cite{Beneke:2000wa} to identify $\xi_\pi(q^2)$ with one of the  
physical form factors. However, for the factorization proof it is more 
convenient to define $\xi_\pi(q^2)$ as a $\SCETI$ matrix element as 
done above.  
 
(ii) If we attempt to factorize $\langle\pi(p')|\bar\xi_C\,W_C \,h_v 
  |\bar B(p)\rangle$ further into light-cone distribution amplitudes,  
we find that the resulting convolution integrals have endpoint 
singularities. Furthermore, three-particle light-cone distribution 
amplitudes of the pion and the $B$ meson would appear already at 
leading power. This can be seen explicitly from the tree-level 
matching of the current to $\SCETII$ in the previous section. We also 
find that $\xi_\pi(q^2)$ scales as $\lambda^3$, since as discussed 
above $\bar\xi_C\,W_C \,h_v$ matches onto $\lambda^8$ operators in  
$\SCETII$.  
 
(iii) From the standpoint of factorization one would like to extract  
the hard-collinear effects from $\xi_\pi(q^2)$ and define a 
non-perturbative ``soft'' form factor, which depends only on 
virtualities of order $\Lambda^2$. In practice there is no gain from 
this, since the main simplification contained in the factorization 
formula is that there is a single form factor that relates all three 
pion form factors, up to a standard hard-scattering term. The 
definition (\ref{xidef}) achieves this simplification. 
 
(iv) Our analysis does not allow conclusions on whether $\xi_\pi(q^2)$ 
is of order $\lambda^3$ or 
$\lambda^3\times\alpha_s(\sqrt{m_B\Lambda})$. Although at least one  
hard-collinear spectator interaction is required to turn the soft 
spectator quark into a collinear quark, the corresponding factor of 
$\alpha_s$ may be compensated by large logarithms. An analysis of 
logarithms is needed that we leave for future investigations. From the 
phenomenological perspective the magnitude of $\xi_\pi(q^2)$ relative 
to the hard-scattering term is  
only a numerical question. 
 
Having defined the $C_i\,\xi_\pi$ term in the factorization formula  
(\ref{fact2}), to complete the factorization proof we must show that the  
matrix element of the second term on the right-hand side of  
(\ref{scet1currents}) factorizes as  
\begin{equation}  
\label{fact3}
\phi_{B} * T_i*\phi_\pi  
\end{equation} 
with convergent convolution integrals of a hard-scattering kernel  
$T_i$ and light-cone distribution amplitudes. 
 
\subsection{Matching on $\SCETII$: Integrating out hard-collinear\\ 
  modes} 
 
We compute the matrix element of the second term on the
right-hand side of (\ref{scet1currents}), 
\begin{equation}
{\cal M}_i \equiv \langle \pi(p')| 
\,\frac{1}{m_b} \ \sum_j \; \int d\hat s_1  
d\hat s_2 \, \tilde C_{ij}^{1\mu}\bigg(\hat s_1, \hat s_2, 
\frac{m_b}{\mu}\bigg) \,  
O_j^{1\mu} (s_1,s_2)|\bar B(p)\rangle
\end{equation}
for $x=0$. Of the three possible Dirac matrices $\Gamma_j^\prime=\{1,
\gamma_5,\gamma_\perp^\nu\}$ only $\gamma_\perp^\nu$ can have a
non-vanishing matrix element, since there is no external transverse 
vector available. Hence we can simplify the previous expression to 
\begin{equation}
\frac{1}{m_b}\int d\hat s_1  
d\hat s_2 \, \frac{g_\perp^{\mu\nu}}{2}\,
\tilde C_{i\nu}^{1\mu}\bigg(\hat s_1, \hat s_2, \frac{m_b}{\mu}\bigg) \,  
 \langle\pi(p')|
(\bar\xi_C \,W_C)_{s_1}(W_C^\dagger i\Slash D_\perp W_C)_{s_2} \,h_v   
|\bar B(p)\rangle.
\end{equation}
Defining the momentum-space coefficient function 
\begin{equation}
 \frac{g_\perp^{\mu\nu}}{2}\,
\tilde C_{i\nu}^{1\mu}\bigg(\hat s_1, \hat s_2, \frac{m_b}{\mu}\bigg) \equiv
\int d\tau_1d\tau \,e^{-i\tau_1 \hat s_1} \,e^{-i\tau (\hat s_2-
\hat s_1)}\,C_i^1\bigg(\tau;\tau_1, \frac{m_b}{\mu}\bigg),
\end{equation}
and proceeding as in the derivation of (\ref{matr1}), we 
obtain
\begin{equation}
{\cal M}_i = \frac{1}{m_b}\int d\tau\,
C_i^1\bigg(\tau;\frac{E}{m_b}, \frac{m_b}{\mu}\bigg) 
\int d\hat s\,e^{-i\tau \hat s}\,
 \langle\pi(p')|
(\bar\xi_C \,W_C)(0)(W_C^\dagger i\Slash D_\perp W_C)(s\np) \,h_v(0)   
|\bar B(p)\rangle.
\label{mm}
\end{equation}
Note that the second integral defines a function $\Xi_\pi(\tau, q^2)$,
so this tells us that before integrating out hard-collinear modes the
three form factors can be expressed in terms of two functions
$\xi_\pi(q^2)$ and  $\Xi_\pi(\tau, q^2)$. Since the second depends
on two variables, this is in general not a useful result, unless the 
three coefficient functions $C_i^1$ are nearly the same, so that the
integral over $\tau$ defines the same function. This will be the 
case in the limit that we neglect hard (but not hard-collinear) 
quantum corrections.

We now show that the operator $O_j^{1\mu} (s_1,s_2)\sim\bar\xi_{C} 
A^\mu_{\perp  hc} h_v$ matches only on four-quark operators of 
the form $(\bar{q}_s h_v) (\bar\xi_c\xi_c)$ with no additional 
fields or derivatives. The most general form a $\SCETII$ operator with 
non-vanishing $\langle\pi|\ldots|\bar B\rangle$ matrix elements can 
take is 
\begin{equation} 
\label{generalop} 
[\mbox{objects}] \times \Big(\bar \xi_c\Gamma_k^\prime h_v\Big) 
\Big(\bar q_s\Gamma_l^\prime\left\{1,\slash n_+/2\right\}\xi_c\Big) .
\end{equation} 
The ``objects'' can be chosen from 
\begin{eqnarray} 
&& n_-^\mu,\, \np^\mu,\, g^{\mu\nu},\, 
\epsilon^{\mu_\perp\nu_\perp\rho\sigma} n_{-\rho} n_{+\sigma},  
\frac{1}{i\nm\partial},\,\frac{1}{i\nm\partial},  
\nonumber\\[0.2cm] 
&&\partial_\perp,\, A_{\perp c},\, A_{\perp s},\, \np\partial,\, \np A_s,\,  
\nm\partial,\, \nm A_c, 
\nonumber\\[0.2cm] 
&& \bar\xi_c \frac{\slash n_+}{2}\Gamma_m^\prime\xi_c, \, 
 \bar q_s \frac{\slash n_+}{2}\Gamma_m^\prime q_s,\, 
 \bar q_s \frac{\slash n_-}{2}\Gamma_m^\prime q_s,\, 
 \bar q_s \Gamma_m^{\prime\prime} q_s, 
\end{eqnarray} 
with $\Gamma_m^{\prime\prime}$ a basis for the remaining eight 
boost-invariant Dirac structures. The notation is symbolic.  
For instance, if the object is 
$1/(i\nm\partial)$, (\ref{generalop}) means that this can act on any 
combination of soft fields that the operator may contain. The 
following power counting argument relies on dimensional analysis and 
boost invariance, so we list the corresponding properties of the 
objects in Table~\ref{tab:count}. Note that we cannot use factors of  
$1/m_b$ or $\nm v$ to build operators, because the hard-collinear 
loops that are eliminated in the matching onto $\SCETII$ do not depend 
on these variables. What makes $\SCETII$ power counting non-trivial is 
the possibility to use $1/(i\nm\partial)$, related to the 
non-localities in the product of soft fields, which in turn is related 
to the fact that the $\nm$ components of soft momenta are as large as 
the $\nm$ components of hard-collinear momenta. Since every factor  
of  $1/(i\nm\partial)$ decreases the 
$\lambda$ scaling of an operator, it is important to constrain 
the number of times this factor can occur.  
 
\begin{table} 
\vspace{0.1cm} 
\begin{center} 
\begin{tabular}{||c|c|c|c|c||c|c|c|c|c||} 
\hline\hline 
&&&&&&&&& \\[-0.7em] 
Object $O$ & $[\lambda]_O$  & boost & $[d]_O$ & $n_i$ & 
Object $O$ & $[\lambda]_O$  & boost & $[d]_O$ & $n_i$  
\\[0.3em]  
\hline\hline 
&&&&&&&&& \\[-0.7em] 
$(i\nm\partial)^{-1} $ & $-2$ & $-1$ & $-1$ & $n_1$ & 
$\bar\xi_c \frac{\slash n_+}{2}\Gamma_m^\prime\xi_c$ & $\phantom{-}4$ & 
$-1$ & $\phantom{-}3$  & $n_8$ \\ 
$(i\np\partial)^{-1} $ & $\phantom{-}0$  & $+1$ & $-1$ &  $n_2$ & 
&&&& \\ 
$\nm^\mu  $ & $\phantom{-}0$  & $+1$ & $\phantom{-}0$ &  $n_3$ & 
$\bar q_s \frac{\slash n_+}{2}\Gamma_m^\prime q_s$ & $\phantom{-}6$& 
$-1$ & $\phantom{-}3$ & $n_{9a}$ \\ 
$\np^\mu  $ & $\phantom{-}0$  & $-1$ & $\phantom{-}0$  &  $n_4$ & 
&&&& \\ 
$\partial_\perp, \,A_{\perp c}, \,A_{\perp s} $ & $\phantom{-}2$  &  
$\phantom{-}0$ & $\phantom{-}1$ & $n_5$ &  
$\bar q_s \frac{\slash n_-}{2}\Gamma_m^\prime q_s$ &  $\phantom{-}6$& 
$+1$ & $\phantom{-}3$ & $n_{9b}$ \\ 
$\np \partial, \,\np A_{s} $ & $\phantom{-}2$  &  
$-1$ & $\phantom{-}1$  & $n_6$ &&&&& \\ 
$\nm \partial,\,\nm A_c $ & $\phantom{-}4$  & $+1$ & $\phantom{-}1$ &
$n_{7}$ & $\bar q_s \Gamma_m^{\prime\prime} q_s$ & $\phantom{-}6$& 
$\phantom{-}0$ & $\phantom{-}3$ &  $n_{9c}$ \\[0.5em]
\hline\hline 
\end{tabular} 
\end{center} 
\centerline{\parbox{14cm}{\caption{\label{tab:count}\small  
Scaling properties of the building blocks of $\SCETII$ operators.  
$[l]_O=n$ means that $O$ scales with $\lambda^n$. The columns labelled 
``boost'' give the scaling $\alpha^n$ of $O$ under boosts $\nm\to\alpha\nm$,  
$\np\to\alpha^{-1}\np$. The mass dimension is denoted $[d]_O$. The 
last column defines the integers $n_i$ that specify the number of 
occurrences of $O$ in an operator.}}} 
\end{table} 
 
Let $n_i$ be the number of times a certain object occurs in an  
operator of the form (\ref{generalop}) as defined in the Table.  
The $\lambda$ scaling $[\lambda]$, boost scaling $[\alpha]$ and  
mass dimension $[d]$ of the operator (\ref{generalop}) are 
then given by 
\begin{eqnarray} 
\label{nn1} 
[\lambda] &=& 10-2 n_1+2 n_5+2 n_6+ 4 n_7+4 n_8+6(n_{9a}+ 
n_{9b}+n_{9c}), 
\\[0.1cm]  
\label{nn2} 
\mbox{[}\alpha] &=& -n_1+n_2+n_3-n_4-n_6+n_{7}-n_8-n_{9a}+n_{9b}, 
\\[0.1cm] 
\label{nn3} 
\mbox{[}d] &=& 6-n_1-n_2+n_5+n_6+n_{7}+3(n_8+n_{9a}+n_{9b}+n_{9c}). 
\end{eqnarray} 
The $\SCETI$ operators we match to $\SCETII$ are boost-invariant, so  
we impose $[\alpha]=0$, and solve (\ref{nn2},\ref{nn3}) for   
$n_1$ and $n_2$ to obtain 
\begin{eqnarray} 
\label{nlam} 
[\lambda] &=& 4+[d]-n_3+n_4+n_5+2 n_6+2 n_{7}+2 n_8+ 4 n_{9a}+ 
2 n_{9b}+3 n_{9c}, 
\\[0.25cm]  
\label{n1} 
n_1 &=& 3-\frac{[d]}{2} +\frac{1}{2} \Big(n_3-n_4+n_5+2 
  n_{7}+2 n_8+2 n_{9a}+4 n_{9b}+3 n_{9c}\Big), 
\\[0.1cm] 
\label{n2} 
n_2 &=& 3-\frac{[d]}{2} +\frac{1}{2} \Big(-n_3+n_4+n_5+2 
  n_6+4 n_8+4 n_{9a}+2 n_{9b}+3 n_{9c}\Big). 
\end{eqnarray} 
Since all the $n_i$ are non-negative, the first equation yields  
$[\lambda] \geq  4+[d]-n_3$. To turn this into something useful, we 
need to limit $n_3$, the number of occurrences of $\nm^\mu$. The 
possible ways that $\nm^\mu$ can appear in (\ref{generalop}) are 
either because there are free Lorentz indices (if the $\SCETI$ 
operator has such indices) or because it multiplies one of the  
$\gamma$ matrices in $\Gamma_k^\prime$, $\Gamma_l^\prime$, 
$\ldots$ The $\SCETI$ operators of interest have only transverse 
indices (from $A_{\perp, hc}^\mu$ or from $\gamma_\perp^\mu$ in  
$\Gamma^\prime_j$), so any factor of $\nm^\mu$ actually contracts to 
0. The one exception is 
$\epsilon^{\mu_\perp \nu_\perp \rho\sigma} n_{-\rho} n_{+\sigma}$.  
Hence $n_3 \leq  n_4$ and  
\begin{equation} 
[\lambda] \geq  4+[d]. 
\end{equation} 
For  $O_j^{1\mu} (s_1,s_2)\sim\bar\xi_C 
A^\mu_{\perp  hc} h_v$ we have $[d]=4$, so $[\lambda]\geq 
8$. Since we are looking for terms of order $\lambda^8$, the only 
solution is $n_5=\ldots n_{9c}=0$, which implies that 
(\ref{generalop}) reduces to a four-quark operator with no additional 
fields or derivatives, and with $n_1=n_2=1$ as was to be shown. 
 
Before we exploit this result it is instructive to reconsider the  
$\SCETI$ operator $\bar\xi_{ hc}\Gamma^\prime_jh_v$, which defines  
$\xi_\pi(q^2)$. Now $[d]=3$, so $[\lambda]\geq 7$. However, $[\lambda]=7$ 
is in fact not possible, since in this case $n_1$ would have to be  
half-integer. The leading contributions have $[\lambda]=8$ and  
either $n_4-n_3=1$, $n_5=\ldots=0$, in which case $n_1=1$, $n_2=2$ or  
$n_3-n_4=0$, $n_5=1$, $n_6=\ldots=0$ in which 
case $n_1=n_2=2$. The second solution corresponds to operators that 
can have an additional transverse soft or collinear gluon field. This 
confirms our earlier conclusion that $\xi_\pi(q^2)$ scales as 
$\lambda^8$, may involve three-particle distributions at leading 
power, and contains divergent convolution integrals (since larger 
values of $n_1$ or $n_2$ lead to more divergent integrals). In
addition we now learn that no four-particle distributions or 
products of three-particle distributions can appear in $\xi_\pi(q^2)$ 
at any order in perturbation theory.

As shown above matching the operator 
$(\bar\xi_C \,W_C)(0)(W_C^\dagger i\Slash D_\perp W_C)(s\np) \,h_v(0) 
\sim \bar\xi_C \Slash A_{\perp hc} h_v$ to $\SCETII$ 
involves only four-quark operators with one 
occurrence of $1/(i\nm\partial)$ and $1/(i\np\partial)$
each at leading order in $\lambda$.
In convolution notation, the operator is 
\begin{eqnarray}
&&\hspace*{-1cm} 
\int_{-\infty}^0 ds'\int^{\infty}_0 dt 
\,\Big[(\bar \xi_c W_c)(x_+)\Gamma_k^\prime (Y_s^\dagger
 h_v)(x_-)\Big]\,
\nonumber\\
&&\hspace*{2cm}\times\,
\Big[(\bar q_s Y_s)(x_-+t\nm)\Gamma_l^\prime\left\{1,\slash
  n_+/2\right\}(W_c^\dagger \xi_c)(x_++s'\np)\Big] 
\label{op4}
\end{eqnarray}
with $x_\mp$ defined in (\ref{xpm}). The operator is local in the
transverse position, but non-local along $\np$ and $\nm$. In the
following we put $x=0$. The solution to (\ref{nlam}) also implied 
$n_3=n_4$. Since $\Gamma_{k,l}^\prime$ do not contain $n_\mp$, 
this excludes the $\slash n_+/2$ structure in the curly 
bracket of (\ref{op4}). This will be important later, because it
follows from this that of the two light-cone distribution amplitudes of the $B$
meson, $\phi_{B\pm}$, only $\phi_{B+}$ appears in the factorization
formula. Furthermore, parity and Lorentz invariance limit the possible
combinations of Dirac matrices to the three cases $\Gamma_k^\prime=
\Gamma_l^\prime$. The colour structure of the operator (\ref{op4}) 
is  $[(\bar\xi_{c}W_c)_a(Y_s^\dagger h_{v})_b]\,
[(\bar q_{s}Y_s)_b(W_c^\dagger\xi_{c})_a]$, because only the 
combination that makes $\bar \xi_c\xi_c$ 
and $\bar q_s h_v$ a colour-singlet will be relevant. The colour
indices $a,b$ will be suppressed.

The convolution integrals in (\ref{op4}) are corrected by
dimensionless coefficient functions, which can depend on dimensionless
ratios of Lorentz invariants. Consider a hard-collinear loop integral
with an insertion of the $\SCETI$ operator and with external legs
corresponding to the four-quark operator above. The momentum space
Feynman rule for the insertion of 
$\int d\hat s\,e^{-i\tau\hat s}\,\bar\xi_C(0)\Slash A_{\perp \rm
  hc}(s\np) h_v(0)$ is $2\pi \delta(\tau-(\nm v)(\np p_3^\prime)/m_b)$, 
where $p_3^\prime$ denotes the momentum of the hard-collinear gluon. 
Therefore only
the transverse and $\nm$ component of the hard-collinear 
$p_3^\prime$ loop integral are
performed, while $\np p_3^\prime$ is kept fixed (and related to
$\tau$ by the delta function). The invariants, on which the result
of a hard-collinear loop integration can depend, are 
$(\np p_j^\prime) (\nm l)$ ($j=1,2,3$) and $\mu^2$, where $p_1^\prime$ and
$p_2^\prime$ are the momenta of the external collinear quark lines,
and $l$ is the momentum of the soft spectator quark. It is important
that the residual momentum of the heavy quark does not appear in this
list. The residual momentum can always be routed through the diagram
such that it flows from the external heavy quark line directly to the
external momentum of the current insertion, so that it never enters
the hard-collinear loop propagators. Trading $p_2^\prime$ for 
$p^\prime$, the independent dimensionless variables can be chosen as 
\begin{equation}
\frac{\np p^\prime\nm l}{\mu^2}, \,\frac{\np p_1^\prime}{\np p^\prime}, \,
\frac{\np p_3^\prime}{\np p^\prime}.
\end{equation}
The coefficient function can depend on powers of logarithms of the
first ratio, but it can in principle be an arbitrary function of the
second two ratios. In terms of the coordinate space convolutions in 
(\ref{op4}) $\nm l$ is replaced by $t^{-1}$, and $\np p_1^\prime$ is 
replaced by $s^{-1}$. Inserting the general coefficient function we
obtain the matching relation 
\begin{eqnarray}
&& \int d\hat s\,e^{-i\tau \hat s}\,
 \langle\pi(p')|
(\bar\xi \,W_c)(0)(W_c^\dagger i\Slash D_\perp W_c)(s\np) \,h_v(0)   
|\bar B(p)\rangle
\nonumber\\
&&\longrightarrow \quad \sum_k
\int ds'\int dt \,
\tilde J_k(m_b\tau/E;\np p^\prime s^\prime, \ln(E/(\mu^2 t)) 
\label{match}\\
&&  
\hspace*{1.2cm}\times 
\langle\pi(p')|\Big[(\bar \xi_c W_c)(0)\Gamma_k^\prime (Y_s^\dagger
 h_v)(0)\Big]\,
\Big[(\bar q_s Y_s)(t\nm)\Gamma_k^\prime(W_c^\dagger \xi_c)(s'\np)\Big] 
|\bar B(p)\rangle
\nonumber
\end{eqnarray}
valid for the leading power form factors. 
We used that the delta function from the Feynman rule 
sets $\np p_3^\prime$ to $m_b\tau/(\nm v)$.

The $\SCETII$ Lagrangian does not couple collinear and soft fields, so
the matrix element formally factorizes 
\begin{eqnarray}
&&\langle\pi(p')|\Big[(\bar \xi_c W_c)(0)\Gamma_k^\prime (Y_s^\dagger
 h_v)(0)\Big]\,
\Big[(\bar q_s Y_s)(t\nm)\Gamma_k^\prime(W_c^\dagger \xi_c)(s'\np)\Big] 
|\bar B(p)\rangle
\nonumber\\[0.2cm]
&& \hspace*{1cm} = 
- (\Gamma_k^\prime)_{\alpha\alpha^\prime}(\Gamma_k^\prime)_{\beta\beta^\prime}
\,\langle\pi(p')|
(\bar \xi_{c\alpha} W_c)(0)(W_c^\dagger \xi_{c \beta^\prime})(s'\np)
|0\rangle 
\nonumber\\[0.2cm]
&&\hspace*{1.5cm}\times\,
\langle 0|(\bar q_{s\beta} Y_s)(t\nm) (Y_s^\dagger h_{v \alpha^\prime})(0) 
|\bar B(p)\rangle.
\label{factorise}
\end{eqnarray}
Potential subtleties are related to the convergence of the convolution
integrals, an issue that we address below. The two matrix elements
define the light-cone distribution amplitudes of the pion and the $B$ 
meson given in (\ref{distamps}). 
These definitions coincide with the standard definition of the 
distribution amplitudes. Inserting them into (\ref{factorise}) we
obtain the two traces 
\begin{equation}
\mbox{tr}\left(\frac{1+\slash v}{2}\slash n_\pm\gamma_5 
\Gamma_k^\prime \frac{\slash n_-}{2}\gamma_5\Gamma_k^\prime \right) 
\tilde \phi_{B\pm}(t)
\end{equation}
multiplied by the position-space 
$B$ meson distribution amplitude. The trace for the 
minus sign vanishes, because $\slash n_-$
commutes or anti-commutes with $\gamma_5\Gamma_k^\prime$, so 
$\tilde \phi_{B-}(t)$ does not appear in the final result. The
momentum space $B$ meson distribution amplitude is defined by 
\begin{equation}
\tilde\phi_{B+}(t)\equiv\int_0^\infty d\omega\,e^{-i\omega t}\,
\phi_{B+}(\omega).
\end{equation}
Using this and (\ref{distamps}) in (\ref{factorise}), 
the right-hand side of (\ref{match}) 
turns into 
\begin{eqnarray}
&&\frac{if_\pi}{4}\,\np p^\prime\,\frac{if_B m_B}{4}\,\nm v\,\sum_k
\mbox{tr}\left(\frac{1+\slash v}{2}\slash n_+\gamma_5 
\Gamma_k^\prime \frac{\slash n_-}{2}\gamma_5\Gamma_k^\prime \right) 
\,\,\int_0^\infty d\omega \int_0^1 du\,
\phi_{B+}(\omega)\,\phi_\pi(u)
\nonumber\\[0.2cm]
&&\hspace*{1.5cm} \times 
\int ds'\int  dt \,e^{-i\omega
  t}\,e^{i(1-u) s^\prime \np p^\prime}
\tilde J_k(m_b\tau/E;\np p^\prime s^\prime, \ln(E/(\mu^2 t)) 
\label{more1}
\end{eqnarray}
The second line defines the momentum-space hard-collinear coefficient
function
\begin{eqnarray}
&&\int ds'\int dt \,e^{-i\omega
  t}\,e^{i(1-u) s^\prime \np p^\prime}
\tilde J_k(m_b\tau/E;\np p^\prime s^\prime, \ln(E/(\mu^2 t)) 
\nonumber\\
&&\equiv \frac{1}{\omega\,\np p'}\, 
J_k(m_b\tau/E; u, \ln(E\omega /\mu^2)).
\label{jcoeff}
\end{eqnarray}
With this definition $J_k$ is a series of powers of $\ln\omega$ given that 
$\tilde J_k$ contained only powers of $\ln t$. Finally, inserting 
(\ref{jcoeff}) into  (\ref{more1}), (\ref{more1}) into  (\ref{match}), and 
(\ref{match}) into (\ref{mm}) we obtain
\begin{eqnarray}
&&\hspace*{-1.5cm}{\cal M}_i \,=\, \frac{1}{m_b}\,
\frac{if_\pi}{4}\,\frac{if_B m_B}{4}\,\nm v\,
\sum_k \mbox{tr}\left(\frac{1+\slash v}{2}\slash n_+\gamma_5 
\Gamma_k^\prime \frac{\slash n_-}{2}\gamma_5\Gamma_k^\prime \right) 
\nonumber\\[0.2cm]
&& \hspace*{-0.8cm} 
\times \int_0^\infty \!d\omega \,\frac{\phi_{B+}(\omega)}{\omega}
\int_0^1 \!du\,\phi_\pi(u)\,
\int \!d\tau\,
C_i^1\bigg(\tau;\frac{E}{m_b}, \frac{m_b}{\mu}\bigg) \,
J_k(m_b\tau/E; u, \ln(E\omega /\mu^2)).
\end{eqnarray}
This has the form of the hard-scattering term in (\ref{fact2}) and 
(\ref{fact3}), where we can now identify the kernels $T_i$ with the
convolutions of a hard coefficient function with the hard-collinear
coefficient functions, $T_i\sim \sum_k C_i^1\star J_k$.

Thus we have shown that the three form factors can be represented 
in terms of s single function $\xi_\pi(q^2)$ plus a hard-scattering
term, which is a convolution of a short-distance kernel 
with the leading twist two-particle light-cone distribution amplitudes
of the pion and the $B$ meson. To complete the proof, we must show
that the convolution integrals converge. The crucial point is that
boost invariance constrained the coefficient functions $J_k$ to 
contain powers of $\ln\omega$ only. The soft convolution integrals 
are therefore the moments
\begin{equation}
\label{softints}
\frac{1}{\lambda_B^{(n)}}\equiv 
\int_0^\infty \!d\omega \,\frac{\phi_{B+}(\omega)}{\omega}\,
\ln^n\!\left(\frac{m_B\omega}{\mu^2}\right), 
\end{equation}
which generalize the moment $\lambda_B=\lambda_B^{(0)}$ introduced in 
\cite{Beneke:1999br}. The integrals converge provided
$\phi_{B+}(\omega)\sim \omega^a$ with $a>0$ for $\omega\to 0$
  and $\phi_{B+}(\omega)\sim \omega^b$ with $b<0$ for 
$\omega\to \infty$. In the Wandzura-Wilczek approximation one finds
that $a=1$ and $\phi_{B+}(\omega)=0$ for $\omega$ larger than a
  critical value \cite{Kawamura:2001jm} implying
  convergence, as do recent QCD sum rule estimates \cite{Ball:2003fq}. 
Furthermore, if the
integrals (\ref{softints}) converge for the distribution amplitude 
evaluated at one scale $\mu_0$, it converges for all scales 
\cite{Lange:2003ff}. It seems safe to conclude that the moments 
(\ref{softints}) exist. This would not be the case for the analogous 
moments of $\phi_{B-}(\omega)$, at least within the 
Wandzura-Wilczek approximation \cite{Beneke:2000wa}, so the absence of
 $\phi_{B-}(\omega)$ from  the factorization formula is crucial.

There is no analogous constraint on the functional form of 
the $u$-integrand, since $J_k$ can be an arbitrary function of 
$u$. However, we argued in Section~\ref{sec:toy} that an 
endpoint singularity of a collinear integral implies that 
soft and collinear contributions must be factorized explicitly, 
so a collinear endpoint divergence should match to a soft endpoint 
divergence to make the sum of all terms unambiguous. But the 
soft convolution integral is convergent, so the $u$-integral must be 
convergent, too.\footnote{A potential collinear endpoint divergence 
could not be related to the $\xi_\pi(q^2)$ term in the 
factorization formula, since with the definition of 
$\xi_\pi(q^2)$ as a matrix element in $\SCETI$ adopted here, 
soft-collinear factorization is never performed for this term.} 
There is no need to be concerned about the convergence of the 
$\tau$-integral. This integral is a convolution 
of the hard matching coefficient with the hard-collinear matching 
coefficient, and originates in the matching of QCD to 
$\SCETI$. If divergent, the integral could be regularized 
perturbatively, and the coefficients $C_i^1$ and 
$J_k$ would be defined accordingly. In any case, we can 
always imagine matching QCD directly to $\SCETII$, in 
which case the integral $\sum _k C_i^1\star \,J_k$ is obtained directly 
as the coefficient function. 
This concludes the proof of the factorization formula. 
 
The arguments presented here carry over to the case of transitions 
to light flavour-non-singlet vector mesons without essential 
modifications. For vector mesons the matrix elements of 
$\bar\xi_C W_C \Gamma_j^\prime \,h_v$ are non-zero for 
$ \Gamma_j^\prime=\{\gamma_5,\gamma_\perp^\mu\}$, so 
instead of (\ref{xidef}) we have two form factors 
commonly denoted as $\xi_\perp(q^2)$ and $\xi_\parallel(q^2)$. 
The conclusion that the $\SCETI$ current operator with 
the hard-collinear gluon field matches to a four-quark operator 
that factorizes into convergent convolutions of light-cone distribution 
amplitudes remains valid, so it follows that the seven independent 
vector meson form factors reduce to two form factors plus the 
hard-scattering term as 
expected \cite{Beneke:2000wa,Charles:1998dr}.

\subsection{Comparison with previous work}

The insights gained from the details of the matching of the
heavy-to-light current to $\SCETII$ and the structure of the 
factorization proof result in a picture that differs in several 
respects from previous work. We briefly mention these 
differences. 

Motivated by the suggestion \cite{Charles:1998dr} that symmetries 
in large-energy effective theory and heavy quark effective theory
reduce the number of independent form factors at large recoil energy, 
the factorization formula (\ref{ff1},\ref{fact2}) was suggested in 
\cite{Beneke:2000wa} and shown to be valid at the level of 
one-gluon exchange. It was noted that formally sub-leading twist-3 
two-particle light-cone distribution amplitudes yield leading-power  
contributions and it was shown that these can be absorbed 
into the definition of the universal form factor(s). The possibility 
that three-particle amplitudes also contribute at leading power 
was not considered. The present work has shown that this is in fact
the case, but again these contributions can be absorbed 
into the definition of the universal form factor(s). 

Subsequent work discussed the $C_i\,\xi_\pi$ term in the factorization
formula in the framework of SCET \cite{Bauer:2000yr} and investigated 
whether some relations between form factors implied by the 
factorization formula could be valid including power 
corrections \cite{BCDF}. These works proceeded on the assumption 
that the leading-power current in $\SCETI$ is 
$\bar\xi_C\hspace*{0.04cm}\Gamma h_v$. From \cite{Bauer:2002aj} and 
the present work we know that formally sub-leading 
currents are also needed. Here we have shown that these 
currents match to conventional hard-scattering contributions 
in $\SCETII$, so that they do not affect the discussion of 
the $C_i\,\xi_\pi$ term. On the other hand, the suppression of 
the matrix element of the formally leading current implies that 
the investigation of power corrections to the leading term is 
difficult, and hence that the conclusions of \cite{BCDF} should be revisited. 
We leave this for the future, but the present work has shown that 
power corrections are suppressed by $1/m_b$ (and not, perhaps, 
$1/\sqrt{m_b}$), a point that was left open in \cite{BCDF}. 

The most detailed discussion of the factorization formula for the 
form factors in the context of SCET is provided 
by \cite{Bauer:2002aj}, where in particular the factorization formula 
has already been claimed to be shown. In addition to mentioning the 
suppression of the leading-power current this paper also suggests 
to prove the factorization formula by two-step matching of QCD to 
$\SCETI$ to $\SCETII$, a line of argument that we followed in the 
present work. However, in our opinion 
the factorization arguments of \cite{Bauer:2002aj} are incomplete or
incorrect in three crucial aspects. All of them have to do with the
fact that the second matching step to $\SCETII$ was not considered 
in sufficient detail. The first point concerns the claim that 
the form factor scales as $\lambda^3$. Since the power counting of 
$\SCETII$ fields and states has not been worked out, the statement
appears to follow from the power counting of the hard-scattering term 
that can be obtained by inspection, and by comparing this to 
the power counting of other non-factorizable terms. Since this 
comparison is done for the time-ordered products in $\SCETI$, but 
without explicit matching to $\SCETII$,   
it is not clear whether all time-ordered products that scale equally 
in $\SCETI$ give equal contributions to the form factors. 
The second point concerns the definition of 
``factorizable'' and ``non-factorizable'' terms. The definition 
in \cite{Bauer:2002aj} is based on the 
formal distinction of $\SCETI$ operators according 
to whether they contain only dressed soft quark fields after the decoupling 
of soft gluons (``factorizable operators'') or also dressed soft 
gluon fields (``non-factorizable''). This should be contrasted with
the point of view taken in this paper that the distinction of 
factorizable and non-factorizable terms is related to soft-collinear 
factorization in $\SCETII$, such that non-factorizable operators are those 
whose matrix elements do not factorize naively into soft and collinear matrix
elements. Comparing the former definition 
to ours, it is not clear to us that 
the term $T_0^F$ identified as ``factorizable'' in 
\cite{Bauer:2002aj} matches only to a four-quark operator without 
endpoint divergences, since the corresponding time-ordered product
defined in $\SCETI$ contains the leading power $\SCETI$ collinear 
interactions that may match to three-particle terms in $\SCETII$. 
Finally, although the paper states that the convolution integrals 
that arise from matching factorizable time-ordered products to
$\SCETII$ are convergent, this claim is not substantiated 
at any point.

\section{Conclusion}
\label{sec:conclude}

To summarize, we have shown that heavy-to-light form factors 
at large recoil energy of the light meson factorize 
according to $F_i = C_i \,\xi_\pi + \phi_B\star T_i\star \phi_\pi$ 
to all orders in $\alpha_s$, and at leading power in an 
expansion in $1/m_b$. The main point of the formula is that 
a larger number of form factors in QCD reduces to a smaller 
number of form factors $\xi_\pi$ (similar to the reduction of the 
number of heavy-to-heavy form factors in heavy quark effective 
theory) up to a hard-scattering term (similar in structure to 
the pion form factor at large momentum transfer). The coefficient 
functions $C_i$ contain hard short-distance effects (scale $m_b$), 
while the $T_i$ represent convolutions of hard and hard-collinear (scale 
$\sqrt{m_B\Lambda}$) effects. All quantities have been defined 
in the framework of soft-collinear effective theory. Because 
the matrix elements of the formally leading currents vanish, the 
factorization formula for the form factors is actually a statement 
about power-suppressed effects.

The more complicated factorization of the heavy-to-light form
factors compared to $B\to D$ form factors and light-meson form factors
is related to the interplay of soft and collinear physics in a 
heavy-to-light transition. This results in multi-particle
contributions and ``endpoint singularities'' at leading power in the
heavy quark expansion, if one attempted to factorize the form factors
along the lines of the pion form factor. In this paper we showed that 
the terms that cause these complications can be absorbed into 
the $C_i\,\xi_\pi$ term and then showed that the remainder factorizes in 
the standard way. This allowed us to by-pass a detailed treatment 
of endpoint singularities. 

We believe, however, that the present
paper provides a first step towards understanding endpoint
singularities in the framework of effective field theory, since 
we identified their origin in the structure of soft-collinear 
factorization within $\SCETII$. While soft and collinear interactions
are formally factorized in the $\SCETII$ Lagrangian, this separation 
requires in general an additional regularization, which 
corresponds to the regularization of endpoint singularities. 
We illustrated how this works for a toy integral, but at 
present it is not clear how to implement the concept of 
soft-collinear factorization in general and in practice. A viable
framework to perform calculations, when endpoint singularities 
cannot be so easily disposed of as in the factorization proof for 
heavy-to-light form factors, is also necessary to understand 
how to sum large logarithms of different scales in the 
hard-scattering kernel $T_i$. We did not address this important 
point here. Finally, we mention that some of our conclusions can 
be adapted to other scattering processes such as the pion form factor 
at sub-leading power, where endpoint singularities are also 
expected to appear. For instance, Figure~\ref{fig:divs} 
has a straightforward translation  to this case, when we 
replace ``soft'' and ``collinear'' by the two types of 
collinear modes (describing the clusters of energetic, nearly massless
particles moving in opposite directions in an appropriate 
reference frame) relevant for light-meson form factors. 
We therefore expect interesting extensions 
of our work into different directions.

\subsubsection*{Acknowledgements}
We would like to thank Thomas Becher, Markus Diehl, 
Enrico Lunghi, Matthias Neubert and many others working in this 
field for interesting discussions. The work of M.B. \ is supported in part 
by the Bundesministerium f\"ur Bildung und Forschung, 
Project 05~HT1PAB/2, and by the DFG Sonder\-forschungsbereich/Transregio~9 
``Computer-ge\-st\"utz\-te Theoretische Teilchenphysik''. 


\end{document}